\documentclass[]{spie}  %>>> use for US letter paper
\usepackage{amsmath,amsfonts,amssymb}
\usepackage{graphicx}
\usepackage[colorlinks=true, allcolors=blue]{hyperref}
\usepackage{tabularx}
\usepackage{booktabs}
\usepackage{caption}
\usepackage{subcaption}
\usepackage{tikz}
\usepackage{float}

\usepackage{lineno}
\usepackage{setspace}
\usepackage{fancyhdr} 

\usepackage[pscoord]{eso-pic}% The zero point of the coordinate systemis the lower left corner of the page (the default).

\newcommand{\placetextbox}[3]{% \placetextbox{<horizontal pos>}{<vertical pos>}{<stuff>}
  \setbox0=\hbox{#3}% Put <stuff> in a box
  \AddToShipoutPictureFG*{% Add <stuff> to current page foreground
    \put(\LenToUnit{#1\paperwidth},\LenToUnit{#2\paperheight}){\vtop{{\null}\makebox[0pt][c]{#3}}}%
  }%
}%

\title{Development of the cadmium zinc TElluride Radiation Imager (TERI)}

\author[a]{Daniel Shy}
\author[b]{Michael Streicher}
\author[b]{Douglas M. Groves}
\author[b]{Zhong He}
\author[b]{Jason Jaworski}
\author[b]{Willy Kaye}
\author[b]{James Mason}
\author[b]{Ryan Parsons}
\author[b]{Feng Zhang}
\author[b]{Yuefeng Zhu}

\author[a]{Alena Thompson}
\author[a]{Alexander Garner}
\author[a]{Anthony Hutcheson}
\author[a]{Mary Johnson-Rambert}
\author[c]{W. Neil  Johnson}
\author[a]{Bernard Phlips}

\usepackage{makecell}

\affil[a]{U.S. Naval Research Laboratory, 4555 Overlook Ave SW, Washington, DC 20375}
\affil[b]{H3D, Inc., 812 Avis Dr., Ann Arbor, MI 48108, USA}
\affil[c]{Technology Service Corporation, Arlington, VA, 22202, USA}

\authorinfo{D. Shy: E-mail: daniel.shy.civ@us.navy.mil}

\begin{document} 
\maketitle

\begin{abstract}

The cadmium zinc TElluride Radiation Imager, or TERI, is an instrument to space qualify large-volume $4 \times 4 \times 1.5 \ \mathrm{cm}^3$ pixelated CdZnTe (CZT) detector technology. The CZT's anode is composed of a $22 \times 22$ array of pixels while the cathode is planar. TERI will contain four of those crystals with each pixel having an energy range of $40 \ \mathrm{keV}$ up to $3 \ \mathrm{MeV}$ with a resolution of $1.3 \%$ full-width-at-half maximum at $662 \  \mathrm{keV}$ all while operating in room temperature. As the detectors are 3D position sensitive, TERI can Compton image events. TERI is fitted with a coded-aperture mask which permits imaging low energy photons in the photoelectric regime. TERI's primary mission is to space-qualify large-volume CZT and measure its degradation due to radiation damage in a space environment. Its secondary mission includes detecting and localizing astrophysical gamma-ray transients. TERI is manifested on DoD's STP-H10 mission for launch to the International Space Station in early 2025. 

\end{abstract}

% Include a list of keywords after the abstract 

\section{INTRODUCTION}
\label{sec:intro}  % \label{} allows reference to this section

\placetextbox{0.5}{0.05}{\large\textsf{DISTRIBUTION STATEMENT A. Approved for public release: distribution is unlimited.}}%

%\placetextbox{0.5}{0.05}{\large\textsf{Not approved for public release.  Further dissemination only by approval of the Commanding Officer, NRL.}}%

To address the gap in astronomical observations of MeV gamma rays, several missions are being developed and conceptualized utilizing various technologies~\cite{MeVAstronomy}. Observations in the MeV range will aid in understanding the formation, evolution, and physics of astrophysical jets, which will identify the physical processes in the extreme conditions around compact objects and the study of the life cycle of matter in the nearby universe. In light of this, we are developing the cadmium zinc TElluride Radiation Imager (TERI) to increase its technological readiness for potential future use of space-based CdZnTe (CZT) sensors.

CZT is by far not new to operating in the space environment. NASA's Swift-BAT Observatory~\cite{swift} has operated planar CZT since 2004 (and is still operational as of February 2024). Coplanar CZT flew on the NASA Dawn spacecraft for an interplanetary mission. Moreover, pixelated CZT with a volume of $4 \times 4 \times 0.5 \ \mathrm{cm}^3$ flew on several Indian Space Research Organization instruments including AstroSat and RT-2~\cite{czti,rt2}. This is the largest crystal we have found that has flown yet. 

We are therefore fielding TERI with four $4 \times 4 \times 1.5 \ \mathrm{cm}^3$ crystals, the largest CZT crystal developed to date~\cite{4x4}. A larger single-volume crystal is desired to result in a higher detection efficiency. A larger single volume reduces the guard-ring-to-volume ratio thereby increasing the detection efficiency~\cite{cztEfficiency}. In addition, larger crystals could also have a positive effect on overall spectroscopic performance. Pixels near the edges generally perform poorly due to the poorer electric field. A larger crystal again reduces the perimeter-to-volume ratio. The improvements originating from larger crystals will enhance the overall efficiency of the imager-spectrometer instrument. 

COSI is a small explorer class telescope in development to survey the soft MeV sky (0.2-5 MeV)~\cite{COSI}. Its technology is based on double-sided double strip detector technology which uses high purity germanium (HPGe) as its detection medium. Due to the nature of HPGe, it needs to be cooled to cryogenic temperatures. This adds complexities to the system, which adds cost and engineering requirements. Room temperature semiconductors, such as CZT, relax the temperature requirements while maintaining good resolution compared to scintillator based instruments. The TERI detectors therefore offer an alternative to HPGe. 

TERI is currently planned for a launch in early 2025 to the International Space Station (ISS) on the Department of Defense (DoD) Space Test Program (STP) STP-H10 mission. It is slated to be hosted on SpaceX Commercial Resupply Services CRS-32. This manuscript serves as its instrument paper. Sec.~\ref{sec:overview} will overview the entire instrument. Sec.~\ref{sec:spectra} described its in-lab spectroscopic performance, Sec.~\ref{sec:ca} describes the coded aperture, while Sec.~\ref{sec:CI} presents its Compton imaging capabilities.

\section{TERI Overview}
\label{sec:overview}

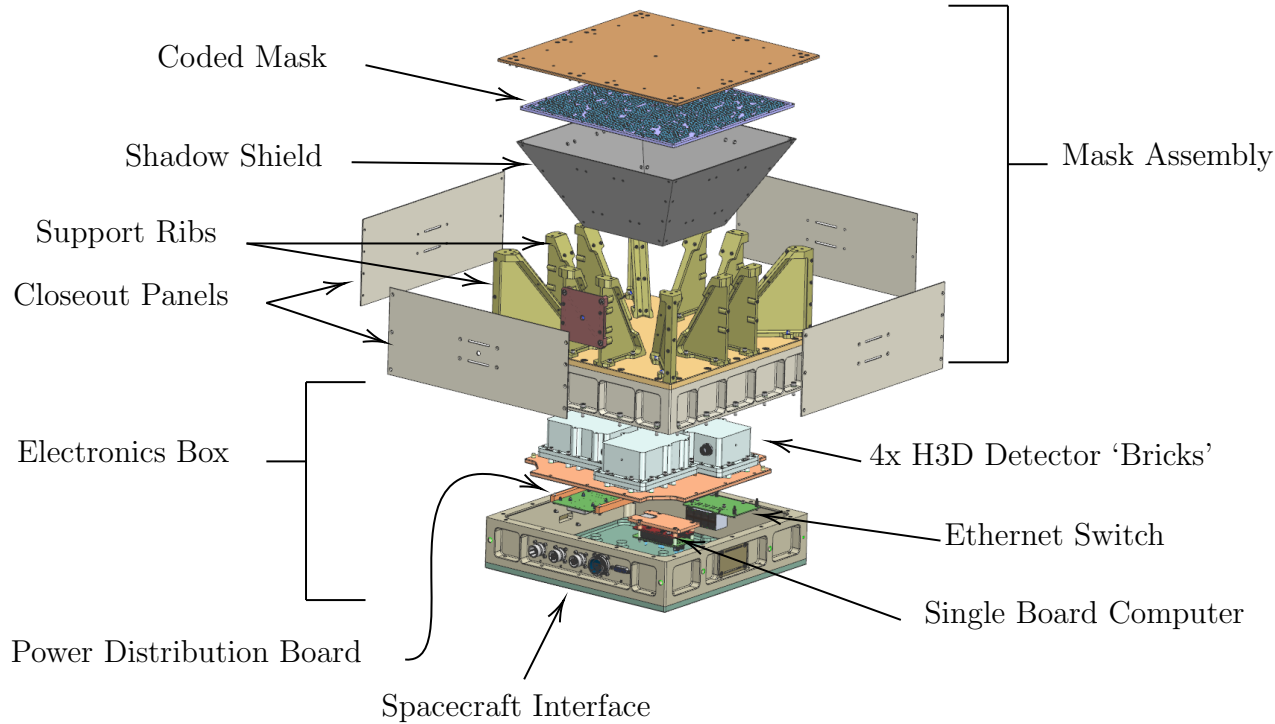
\begin{figure}[h]

\begin{center}

\tikzset{every picture/.style={line width=0.75pt}} %set default line width to 0.75pt        

\begin{tikzpicture}[x=0.75pt,y=0.75pt,yscale=-1,xscale=1]
%uncomment if require: \path (0,391); %set diagram left start at 0, and has height of 391

%Image [id:dp8460490973106158] 
\draw (342,160.06) node  {\includegraphics[width=237pt,height=238.13pt]{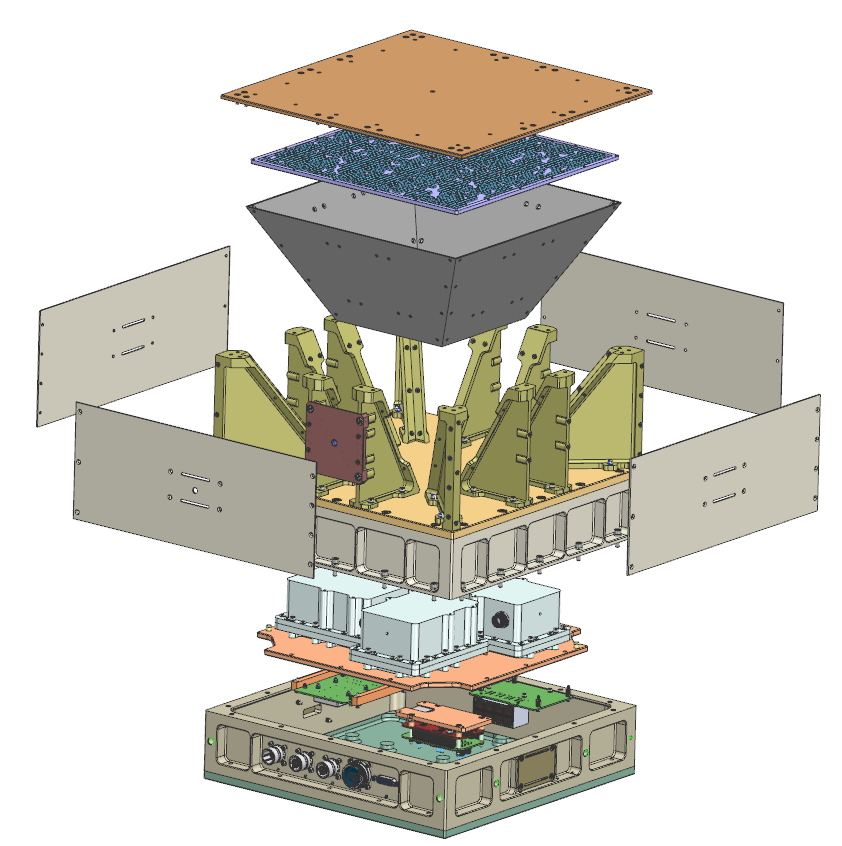}};
%Straight Lines [id:da29404491688666023] 
\draw    (494,10) -- (524,10) ;
%Straight Lines [id:da14864565726484102] 
\draw    (494,190) -- (524,190) ;
%Straight Lines [id:da1500226512102215] 
\draw    (524,10) -- (524,190) ;
%Straight Lines [id:da7535159104644076] 
\draw    (524,90) -- (544,90) ;
%Straight Lines [id:da4035505419547696] 
\draw    (170,310) -- (200,310) ;
%Straight Lines [id:da27339239264197734] 
\draw    (170,200) -- (200,200) ;
%Straight Lines [id:da6146291306234503] 
\draw    (170,200) -- (170,310) ;
%Straight Lines [id:da29320711885292305] 
\draw    (150,240) -- (170,240) ;
%Straight Lines [id:da16777735356331225] 
\draw    (200,40) -- (275.05,56.57) ;
\draw [shift={(277,57)}, rotate = 192.45] [color={rgb, 255:red, 0; green, 0; blue, 0 }  ][line width=0.75]    (10.93,-3.29) .. controls (6.95,-1.4) and (3.31,-0.3) .. (0,0) .. controls (3.31,0.3) and (6.95,1.4) .. (10.93,3.29)   ;
%Straight Lines [id:da6794447817347551] 
\draw    (193,89) -- (278,89.98) ;
\draw [shift={(280,90)}, rotate = 180.66] [color={rgb, 255:red, 0; green, 0; blue, 0 }  ][line width=0.75]    (10.93,-3.29) .. controls (6.95,-1.4) and (3.31,-0.3) .. (0,0) .. controls (3.31,0.3) and (6.95,1.4) .. (10.93,3.29)   ;
%Straight Lines [id:da06170594760865733] 
\draw    (150,160) -- (188.06,150.49) ;
\draw [shift={(190,150)}, rotate = 165.96] [color={rgb, 255:red, 0; green, 0; blue, 0 }  ][line width=0.75]    (10.93,-3.29) .. controls (6.95,-1.4) and (3.31,-0.3) .. (0,0) .. controls (3.31,0.3) and (6.95,1.4) .. (10.93,3.29)   ;
%Straight Lines [id:da6989169768161412] 
\draw    (150,160) -- (208.1,179.37) ;
\draw [shift={(210,180)}, rotate = 198.43] [color={rgb, 255:red, 0; green, 0; blue, 0 }  ][line width=0.75]    (10.93,-3.29) .. controls (6.95,-1.4) and (3.31,-0.3) .. (0,0) .. controls (3.31,0.3) and (6.95,1.4) .. (10.93,3.29)   ;
%Straight Lines [id:da8266482242245156] 
\draw    (140,130) -- (288,130) ;
\draw [shift={(290,130)}, rotate = 180] [color={rgb, 255:red, 0; green, 0; blue, 0 }  ][line width=0.75]    (10.93,-3.29) .. controls (6.95,-1.4) and (3.31,-0.3) .. (0,0) .. controls (3.31,0.3) and (6.95,1.4) .. (10.93,3.29)   ;
%Straight Lines [id:da08240667692339121] 
\draw    (140,130) -- (258.03,149.67) ;
\draw [shift={(260,150)}, rotate = 189.46] [color={rgb, 255:red, 0; green, 0; blue, 0 }  ][line width=0.75]    (10.93,-3.29) .. controls (6.95,-1.4) and (3.31,-0.3) .. (0,0) .. controls (3.31,0.3) and (6.95,1.4) .. (10.93,3.29)   ;
%Straight Lines [id:da9235909186407184] 
\draw    (490,280) -- (400.98,267.28) ;
\draw [shift={(399,267)}, rotate = 8.13] [color={rgb, 255:red, 0; green, 0; blue, 0 }  ][line width=0.75]    (10.93,-3.29) .. controls (6.95,-1.4) and (3.31,-0.3) .. (0,0) .. controls (3.31,0.3) and (6.95,1.4) .. (10.93,3.29)   ;
%Straight Lines [id:da7726908246861225] 
\draw    (450,240) -- (405.95,230.42) ;
\draw [shift={(404,230)}, rotate = 12.26] [color={rgb, 255:red, 0; green, 0; blue, 0 }  ][line width=0.75]    (10.93,-3.29) .. controls (6.95,-1.4) and (3.31,-0.3) .. (0,0) .. controls (3.31,0.3) and (6.95,1.4) .. (10.93,3.29)   ;
%Straight Lines [id:da7725824854184383] 
\draw    (470,320) -- (361.88,280.68) ;
\draw [shift={(360,280)}, rotate = 19.98] [color={rgb, 255:red, 0; green, 0; blue, 0 }  ][line width=0.75]    (10.93,-3.29) .. controls (6.95,-1.4) and (3.31,-0.3) .. (0,0) .. controls (3.31,0.3) and (6.95,1.4) .. (10.93,3.29)   ;
%Straight Lines [id:da5159094437565593] 
\draw    (280,350) -- (299.11,311.79) ;
\draw [shift={(300,310)}, rotate = 116.57] [color={rgb, 255:red, 0; green, 0; blue, 0 }  ][line width=0.75]    (10.93,-3.29) .. controls (6.95,-1.4) and (3.31,-0.3) .. (0,0) .. controls (3.31,0.3) and (6.95,1.4) .. (10.93,3.29)   ;
%Curve Lines [id:da22442345525896035] 
\draw    (220,340) .. controls (255.65,343.07) and (197.42,215.37) .. (287.63,253.41) ;
\draw [shift={(289,254)}, rotate = 203.42] [color={rgb, 255:red, 0; green, 0; blue, 0 }  ][line width=0.75]    (10.93,-3.29) .. controls (6.95,-1.4) and (3.31,-0.3) .. (0,0) .. controls (3.31,0.3) and (6.95,1.4) .. (10.93,3.29)   ;

% Text Node
\draw (143.5,35.5) node  [font=\large] [align=left] {\begin{minipage}[lt]{72.76pt}\setlength\topsep{0pt}
{\fontfamily{cmr}\selectfont Coded Mask}
\end{minipage}};
% Text Node
\draw (136.5,85.5) node  [font=\large] [align=left] {\begin{minipage}[lt]{86.64pt}\setlength\topsep{0pt}
{\fontfamily{cmr}\selectfont Shadow Shield}
\end{minipage}};
% Text Node
\draw (83,155.5) node  [font=\large] [align=left] {\begin{minipage}[lt]{91.12pt}\setlength\topsep{0pt}
{\fontfamily{cmr}\selectfont Closeout Panels}
\end{minipage}};
% Text Node
\draw (630,86.5) node  [font=\large] [align=left] {\begin{minipage}[lt]{118.32pt}\setlength\topsep{0pt}
{\fontfamily{cmr}\selectfont Mask Assembly}
\end{minipage}};
% Text Node
\draw (85,235) node  [font=\large] [align=left] {\begin{minipage}[lt]{91.12pt}\setlength\topsep{0pt}
{\fontfamily{cmr}\selectfont Electronics Box}
\end{minipage}};
% Text Node
\draw (572,276.5) node  [font=\large] [align=left] {\begin{minipage}[lt]{120.02pt}\setlength\topsep{0pt}
{\fontfamily{cmr}\selectfont  Ethernet Switch}
\end{minipage}};
% Text Node
\draw (580.5,317.5) node  [font=\large] [align=left] {\begin{minipage}[lt]{147.97pt}\setlength\topsep{0pt}
{\fontfamily{cmr}\selectfont Single Board Computer}
\end{minipage}};
% Text Node
\draw (130,335.5) node  [font=\large] [align=left] {\begin{minipage}[lt]{163.2pt}\setlength\topsep{0pt}
{\fontfamily{cmr}\selectfont Power Distribution Board}
\end{minipage}};
% Text Node
\draw (91.5,125.5) node  [font=\large] [align=left] {\begin{minipage}[lt]{86.64pt}\setlength\topsep{0pt}
{\fontfamily{cmr}\selectfont Support Ribs}
\end{minipage}};
% Text Node
\draw (562,236.5) node  [font=\large] [align=left] {\begin{minipage}[lt]{161.84pt}\setlength\topsep{0pt}
{\fontfamily{cmr}\selectfont  4x H3D Detector `Bricks'}
\end{minipage}};
% Text Node
\draw (287,364.5) node  [font=\large] [align=left] {\begin{minipage}[lt]{118.32pt}\setlength\topsep{0pt}
{\fontfamily{cmr}\selectfont Spacecraft Interface }
\end{minipage}};

\end{tikzpicture}

\end{center}

\caption{Exploded view of the TERI (computer aided design model). The mechanical interface to the Columbus CEPA pallet is not depicted.}
\label{fig:teriExploded}
\end{figure}

\begin{figure}
     \centering
     \begin{subfigure}[b]{0.45\textwidth}
         \centering
         \subcaption{Cross-Sectional View}
         \includegraphics[trim={7cm 1cm 3cm 1cm}, clip, height=0.97\textwidth]{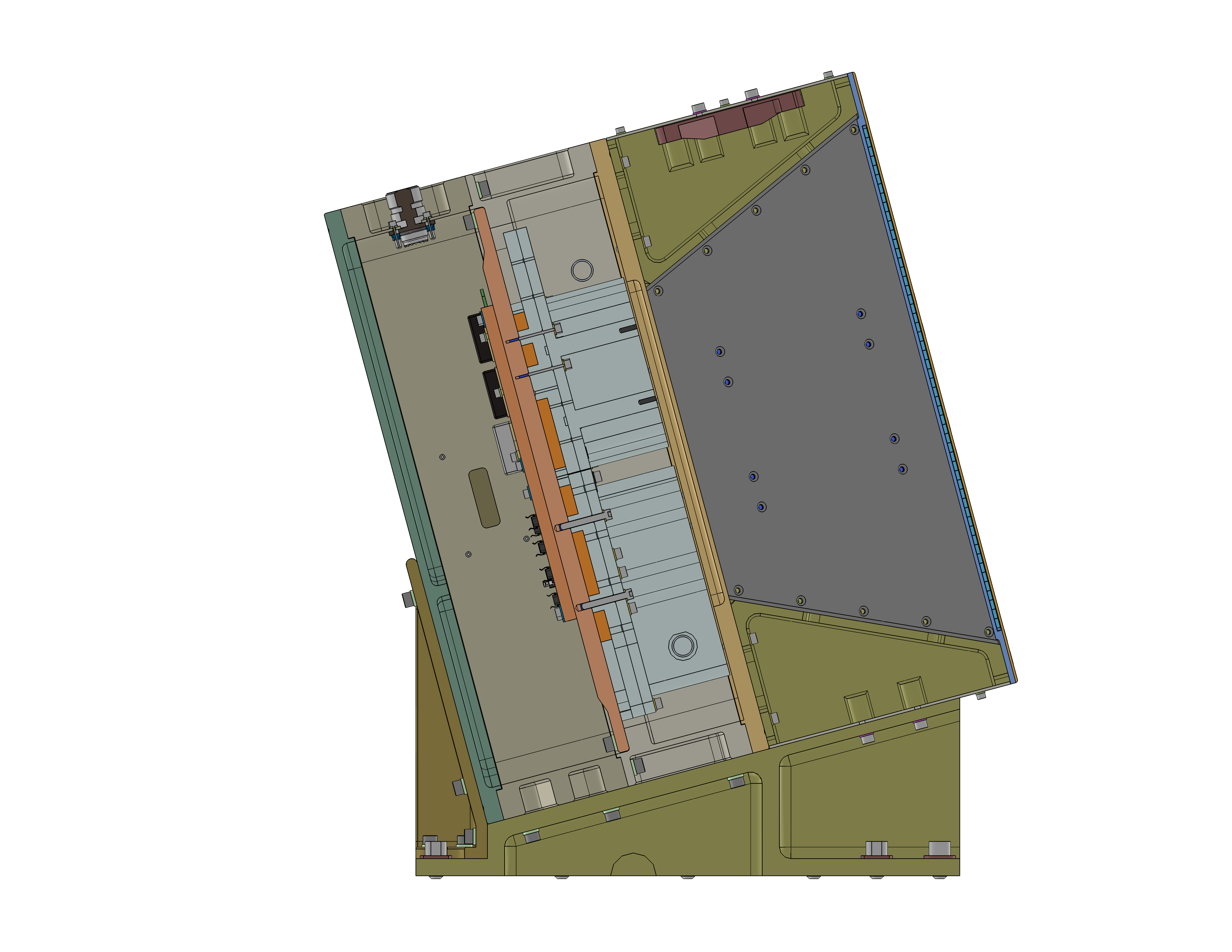}
         \label{fig:TERI_CrossSection}
     \end{subfigure}
     \hspace{15mm}
     \begin{subfigure}[b]{0.45\textwidth}
         \centering
           \caption{TERI Picture}
           \rotatebox[origin=c]{90}{\includegraphics[trim={5cm 0cm 2cm 1cm}, clip, width=1\textwidth]{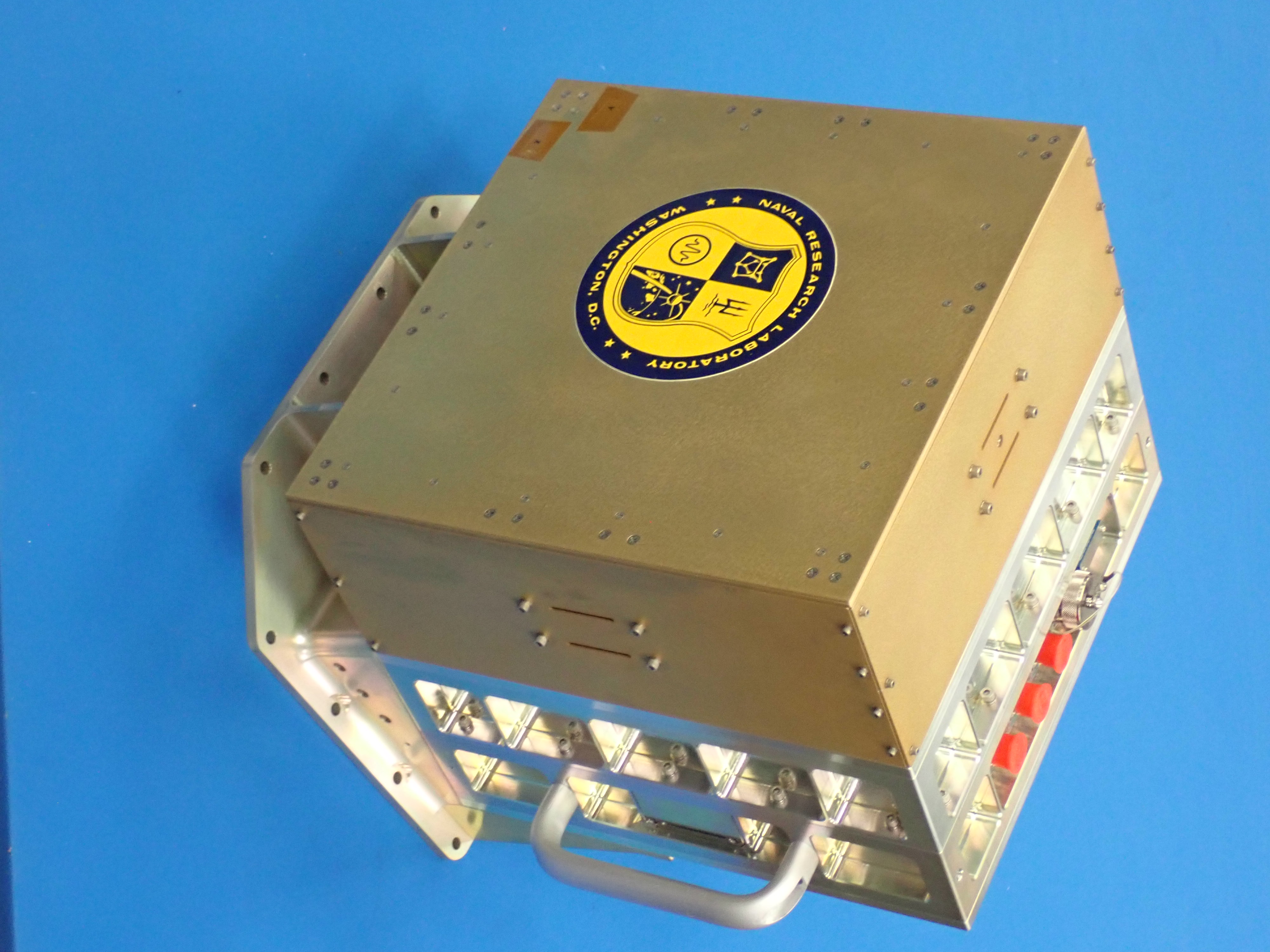}}
         \label{fig:teriFull}
     \end{subfigure}

        \caption{Figures of TERI. (a) Shows a cross-sectional view with the Zenith's vector roughly pointing to the right. The ISS forward direction is out of the page. (b) Picture of the fully assembled TERI.}
        \label{fig:teriPics}
\end{figure}

The cadmium zinc TElluride Radiation Imager, or TERI, has a total volume of $35 \mathrm{(L)} \times 35 \mathrm{(W)} \times 31 \mathrm{(H)} \ \mathrm{cm}^3$. It weighs $\sim 40 \ \mathrm{kg}$ and has a power draw of $\sim30 \ \mathrm{W}$. Fig.~\ref{fig:teriExploded} shows an exploded view of TERI. There are two major components, the electronics housing and the mask assembly. The electronics housing contains the 4 CZT detectors, a power distribution unit, a single board computer (Raspberry Pi) to act as the master flight computer, and an Ethernet switch. The upper half of the instrument consists of the coded mask assembly, which is discussed in Sec.~\ref{sec:ca}. Fig.~\ref{fig:TERI_CrossSection} shows a cross-sectional view of TERI while Fig.~\ref{fig:teriFull} shows the fully assembled instrument.

\begin{table}[H]
  \centering
  \caption{Table with TERI's instrument performance. The unshielded field of view (FOV) refers to the opening of the shadow shield. The Compton imaging resolution is calculated with a $\mathrm{Cs}$-137 source placed in the center of a single TERI detector. The reported value reflects the FWHM of the reconstructed point spread function (PSF). The Maximum Likelihood Expectation Maximization (MLEM) value is calculated under similar conditions using 25 MLEM iterations. See Section~\ref{sec:CI} for additional information on TERI's Compton imaging capabilities.}
	\begin{tabular}{c|c}	
	    \toprule
		Metric	& Value  \\ \hline \hline
		Energy Range 	& $40 \ \mathrm{keV} - 3 \ \mathrm{MeV}$ per pixel \\ \hline
		Energy Resolution & $1.3\%  \ \mathrm{FWHM} \ @ \ 662 \ \mathrm{keV}$ \\ \hline
		Unshielded FOV & $0.69 \ \mathrm{sr}$ \\ \hline
		Compton Imaging Resolution & \makecell{$45^\circ \ \mathrm{FWHM \ (SBP)}$ \\ $<12^\circ \ \mathrm{FWHM \ (MLEM-25)}$} \\ \hline
		\makecell{Coded Aperture \\ Imaging Resolution}  & $<2^\circ \ \mathrm{FWHM}$ \\ \hline
		Volume & $35 \mathrm{(L)} \times 35 \mathrm{(W)} \times 31 \mathrm{(H)} \ \mathrm{cm}^3$ \\ \hline
		Mass & $\sim 40 \ \mathrm{kg}$ \\ \hline
		Power & $\sim30 \ \mathrm{W}$ \\
		\bottomrule		 
	\end{tabular} 
	\label{tab:teriCharacteristics}
\end{table}

\subsection{Mission Programmatics}

Launch is provided by the Space Test Program (STP), under the United States Space Force, with mission support for at least one year. TERI is one of a few instruments integrated on a Columbus External Payload Adapter (CEPA) pallet designated STP-H10. The pallet's avionics supplies 28V to the instrument along with a GPS-synchronized
pulse per second (PPS) signal. Communication between the pallet and TERI is performed via an asynchronous RS-422. The mission's payload is slated to occupy Starboard Overhead X (SOX) pointing spot on the \textit{Columbus} external payload facility. TERI will be pointed $15^{\circ}$ away from the ISS overhead vector towards the ISS starboard to prevent field-of-view obstruction from other ISS structures and visiting vehicles. TERI's orientation is fixed relative to the ISS and will not alter during flight.

\begin{figure}[h]
  \centering
  \includegraphics[trim={0cm 0cm 8cm 0cm}, clip, width=0.6\linewidth]{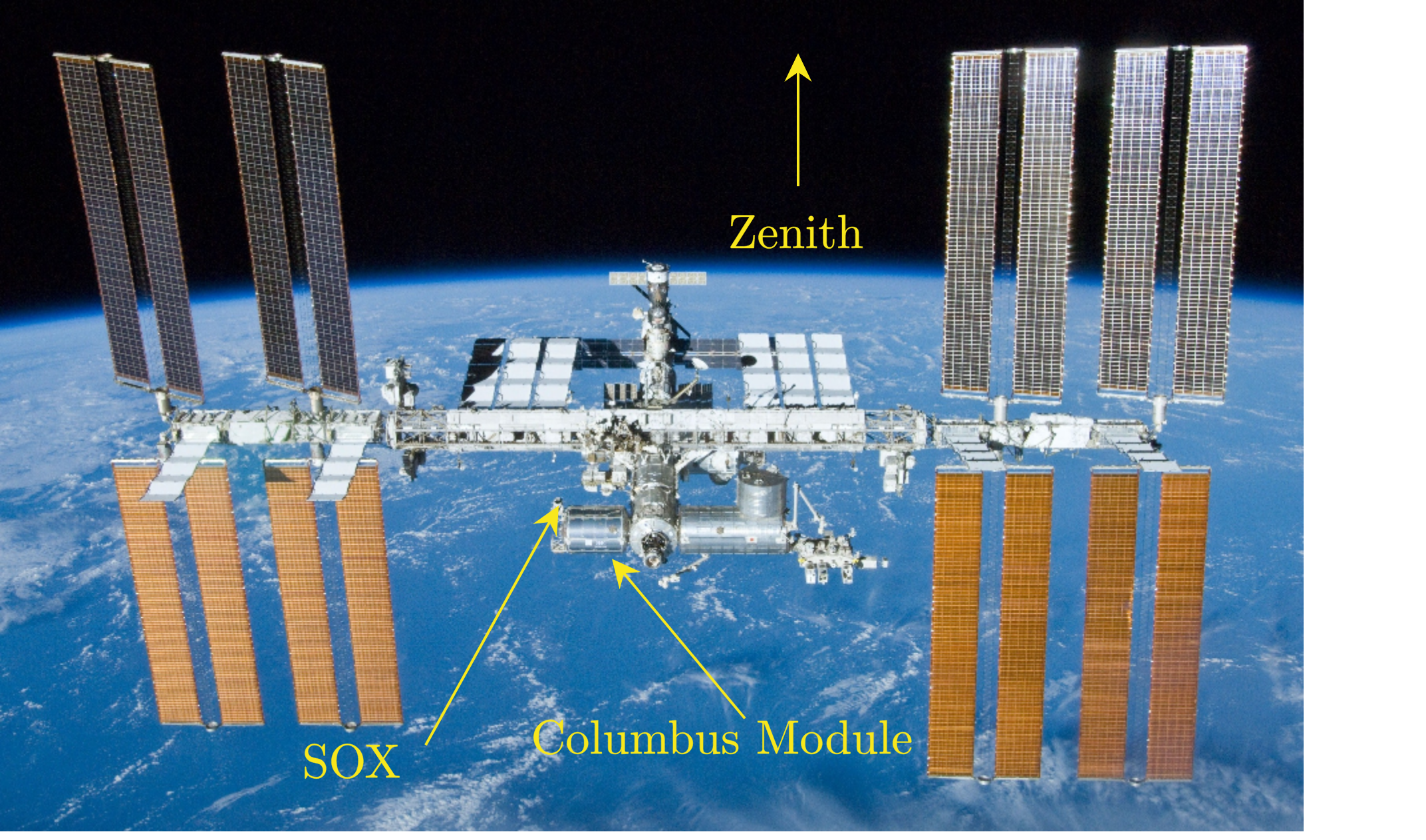}
  \caption{Picture of the International Space Station with SOX and the Columbus module identified. Image is adapted from NASA.gov~\cite{issPic}.}
  \label{fig:iss}
\end{figure}

\subsection{TERI's CZT Detectors}

Each CZT crystal has a volume of $4 \times 4 \times 1.5 \ \mathrm{cm}^3$. The anode is pixelated with a $22 \times 22$ array and a pixel pitch of $1.77 \ \mathrm{mm}$, while the cathode remains planar. This electrode configuration allows for the 3D position estimation of each interaction. We voxelized the interactions using the 2D location of the triggered pixel's center. The 3rd dimension, or the depth of interaction, is estimated using one of two ways. If a single interaction in the crystal has occurred, we use the signals from the cathode and anode and take their ratio (cathode-to-anode ratio). We will use the drift time to estimate the depth if more than one interaction has occurred in the crystal. The techniques for the depth of interaction estimation are extensively explored in He~\cite{HeRamoReview}, Kaye~\cite{KayeThesis}, and Zhang~\cite{Feng3DCZT}. We then apply a voxel by voxel energy correction~\cite{3dczt}.

\begin{figure}
     \centering
     \begin{subfigure}[b]{0.45\textwidth}
         \centering
         \subcaption{$4 \times 4 \times 1.5 \ \mathrm{cm}^3$ CdZnTe}
         \includegraphics[trim={20cm 15cm 30cm 25cm}, clip,height=0.6\textwidth]{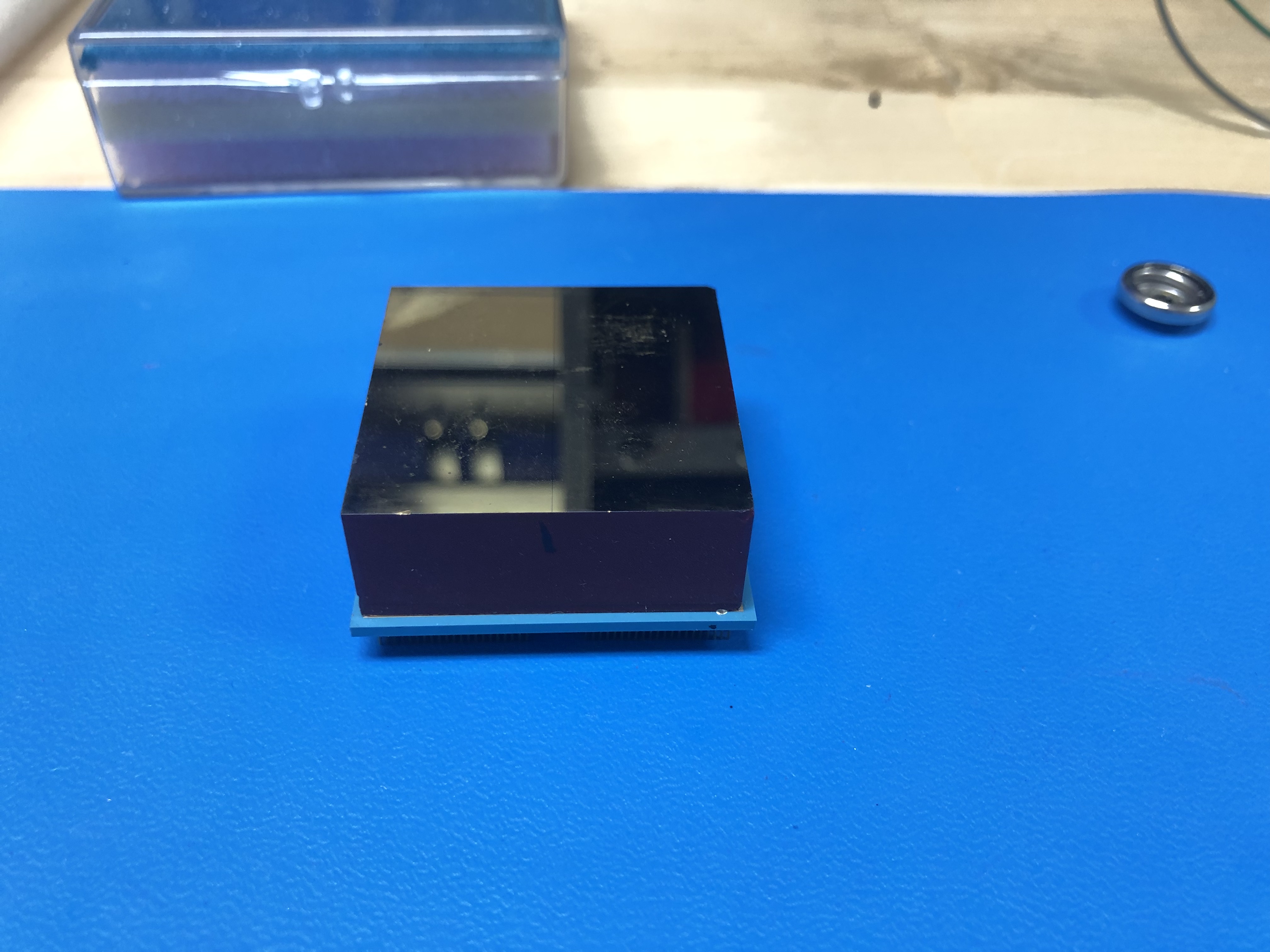}
         \label{fig:crystal}
     \end{subfigure}
     \hspace{0.1mm}
     \begin{subfigure}[b]{0.45\textwidth}
         \centering
         \caption{Electronics housing}
         \includegraphics[height=0.6\textwidth]{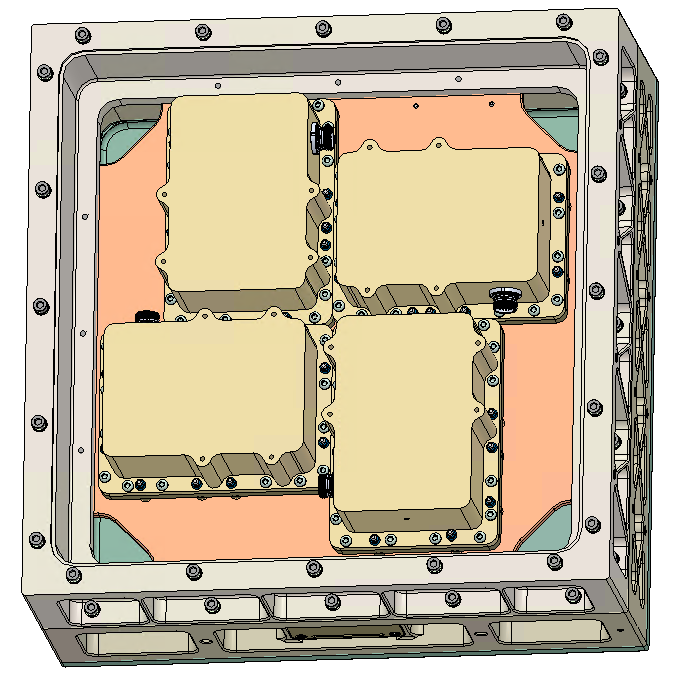}
         \label{fig:electricalBox}
     \end{subfigure}
        \caption{Figures showing the detectors in TERI. (a) Shows the $4 \times 4 \times 1.5 \ \mathrm{cm}^3$ CdZnTe. (b) Shows a CAD of the top view of the electronics box showing the 4 detector modules in a pinwheel design. The dimensions of the electronics box is $35 \mathrm{(L)} \times 35 \mathrm{(W)} \times 16.2 \mathrm{(H)} \ \mathrm{cm}^3$.}
        \label{fig:detectors}
\end{figure}

Fig.~\ref{fig:crystal} shows one of the flight CZT crystals. The rest of the detector unit, which we will refer to as a module is manufactured by H3D, Inc., and is a customized version of the T400 model. Due to the nature of the detector readout, coincidence cannot be accomplished between multiple detectors.

The modules are placed in a pinwheel design to keep a compact form factor while maintaining the crystals square to each other, as visible in Fig.~\ref{fig:electricalBox}. The center-to-center distance between each crystal is $~9.9 \ \mathrm{cm}$.

\begin{table}[h!]
  \centering
  \caption{Table with the detector characteristics.}
	\begin{tabular}{c|c}	
	    \toprule
		Metric	& Value  \\ \hline \hline
		CZT Volume 	& $4 \times 4 \times 1.5 \ \mathrm{cm}^3$ \\ \hline
		Pixel Configuration & $22 \times 22$ array\\ \hline
		Pixel Pitch & $1.77 \ \mathrm{mm}$ \\ \hline
		\bottomrule		 
	\end{tabular} 
	\label{tab:cztCharacteristics}
\end{table}

\subsection{Data and Power}

\begin{figure}[h]
  \centering
  \includegraphics[trim={0cm 0cm 0cm 0cm}, clip, width=0.7\linewidth]{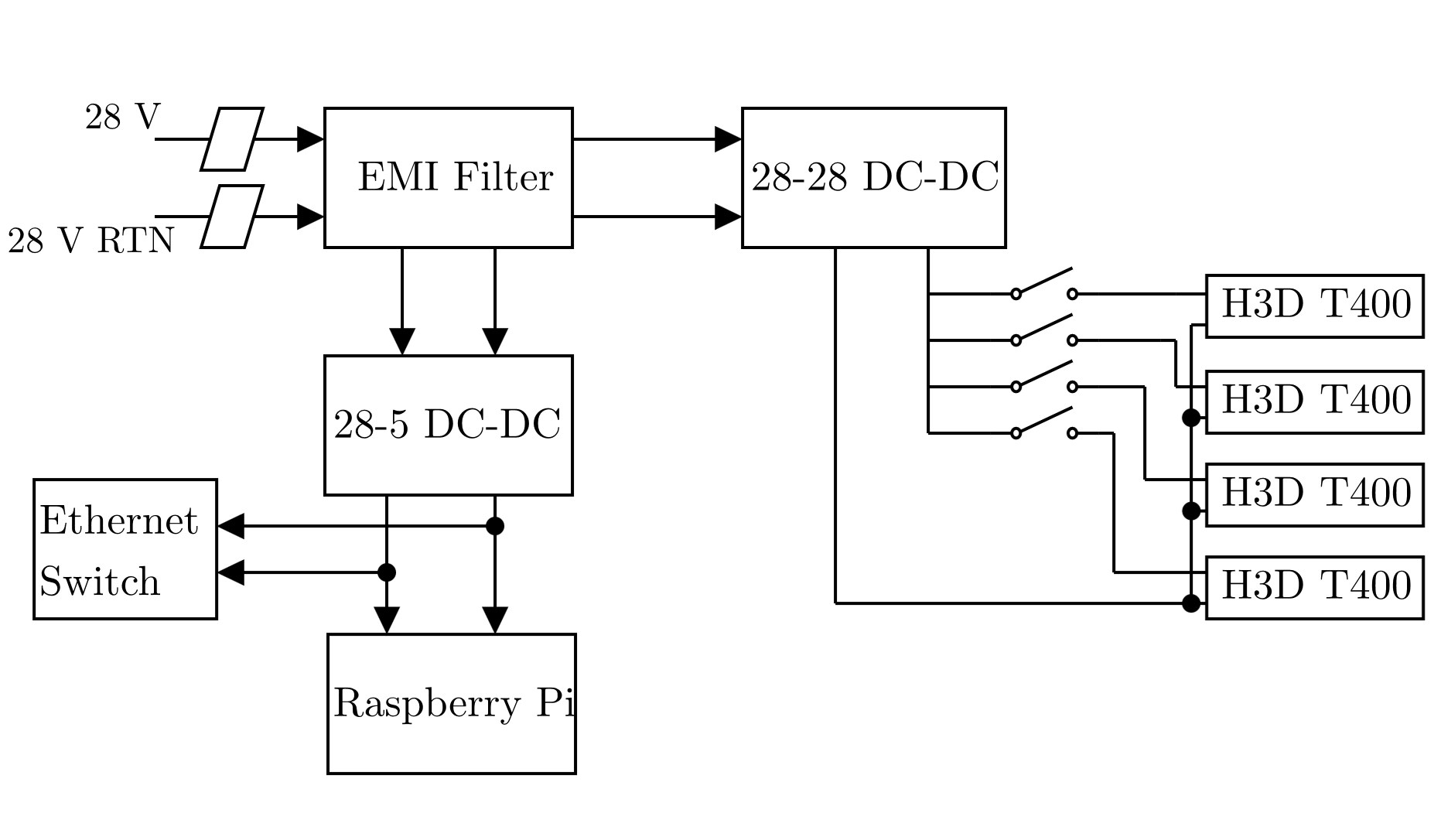}
  \caption{Electrical schematic of the TERI system. The pallet avionics tower provides 28V power.}
  \label{fig:EE_Diagram}
\end{figure}

Fig.~\ref{fig:EE_Diagram} shows the basic electrical schematic of TERI. The 28V line is provided by the pallet's avionics tower. It is then filtered, and split into a 28V DC-DC converter and a 5V DC-DC converter. The 28V provides power to the detector modules. The 5V line provides power to a Raspberry Pi (the flight computer) and an Ethernet switch. The Raspberry Pi communicates with each of the detectors via the Ethernet switch. Each detector provides the 3D position of each interaction along with the reconstructed energy and time.  It monitors voltages, currents, and temperature probes located within TERI. The housekeeping data along with histogramed and list-mode data from the detectors is telemetered to the ground with the Consultative Committee for Space Data Systems (CCSDS) protocol.

\section{Spectroscopic Performance}
\label{sec:spectra}

\begin{figure}[H]
  \centering
  \includegraphics[trim={0cm 0cm 0cm 0cm}, clip, width=0.7\linewidth]{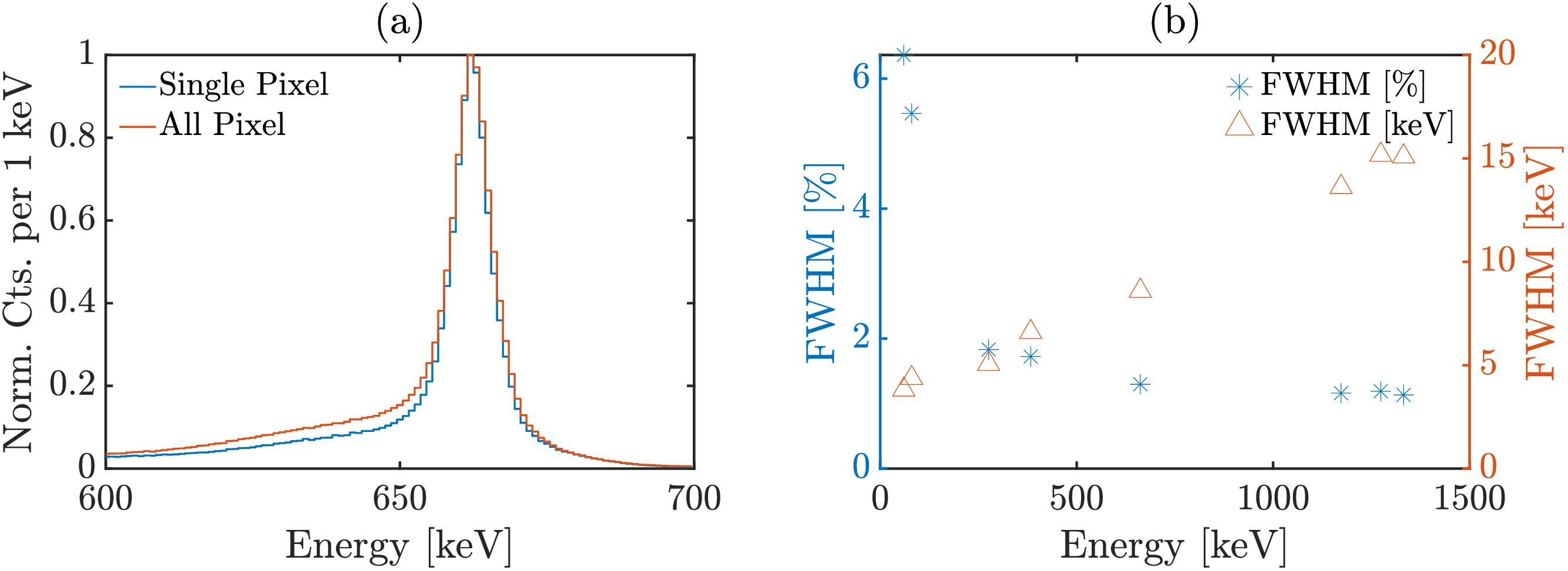}
  \caption{(a) Cs-137 spectra demonstrating the difference in resolution between single pixel and all events. (b) FWHM as a function of energy using all events. }
  \label{fig:resolution}
\end{figure}

The general performance of the $4 \times 4 \times 1.5 \  \mathrm{cm}^3$ CZT crystals is extensively reported in Zhu et alli~\cite{4x4}. A 3D voxel-by-voxel calibration is applied for energy reconstruction that is also temperature dependent~\cite{HE1999173}. The measured full width at half maximum (FWHM) at $662 \ \mathrm{keV}$ is $1.3 \%$ using all events and $1.2 \%$ using single-pixel events. Fig~\ref{fig:resolution}a plots the $662 \ \mathrm{keV}$ full energy peak using either single pixel or all pixel events. The resolution, as expected, is better for single-pixel events. Fig.~\ref{fig:resolution}b plots the FWHM as a function of energy. Note that this plot uses all pixel event types.

\begin{figure}[H]
  \centering
  \includegraphics[trim={0cm 0cm 0cm 0cm}, clip, width=0.7\linewidth]{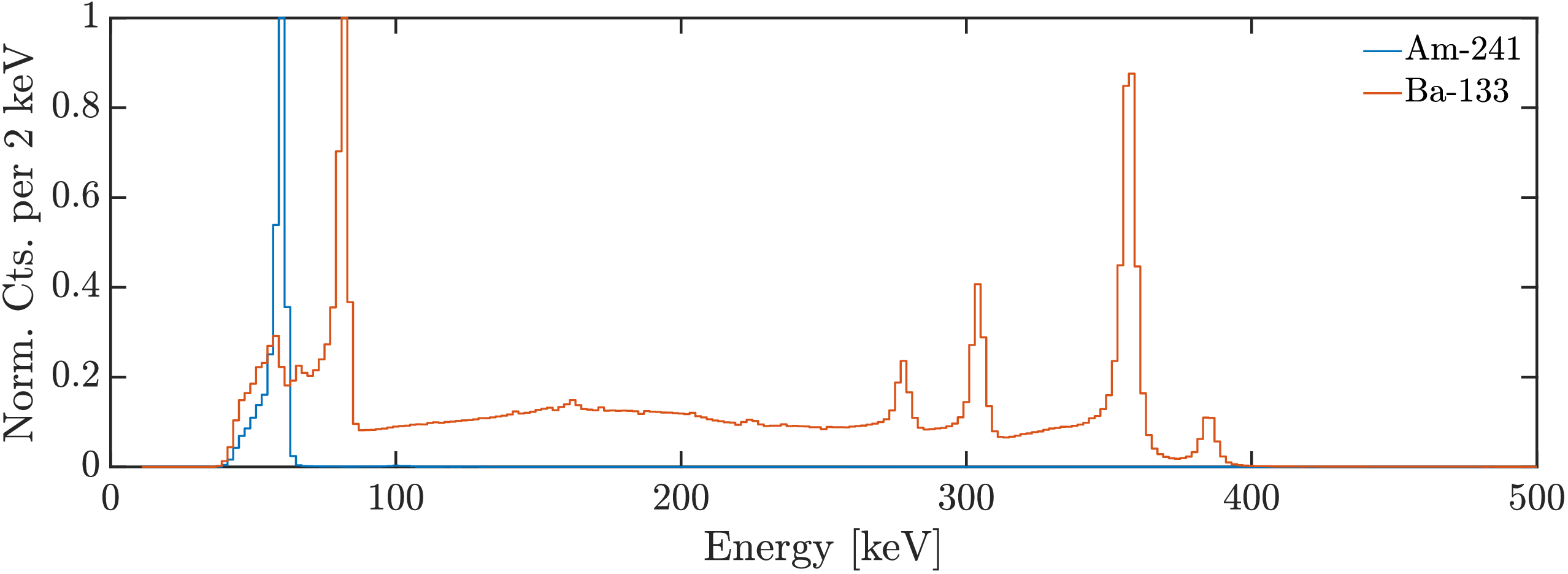}
  \caption{Energy spectra of Am-241 and Ba-133 taken by TERI. All pixel events are plotted.}
  \label{fig:lowErgSpectra}
\end{figure}

\begin{figure}[H]
  \centering
  \includegraphics[trim={0cm 0cm 0cm 0cm}, clip, width=0.7\linewidth]{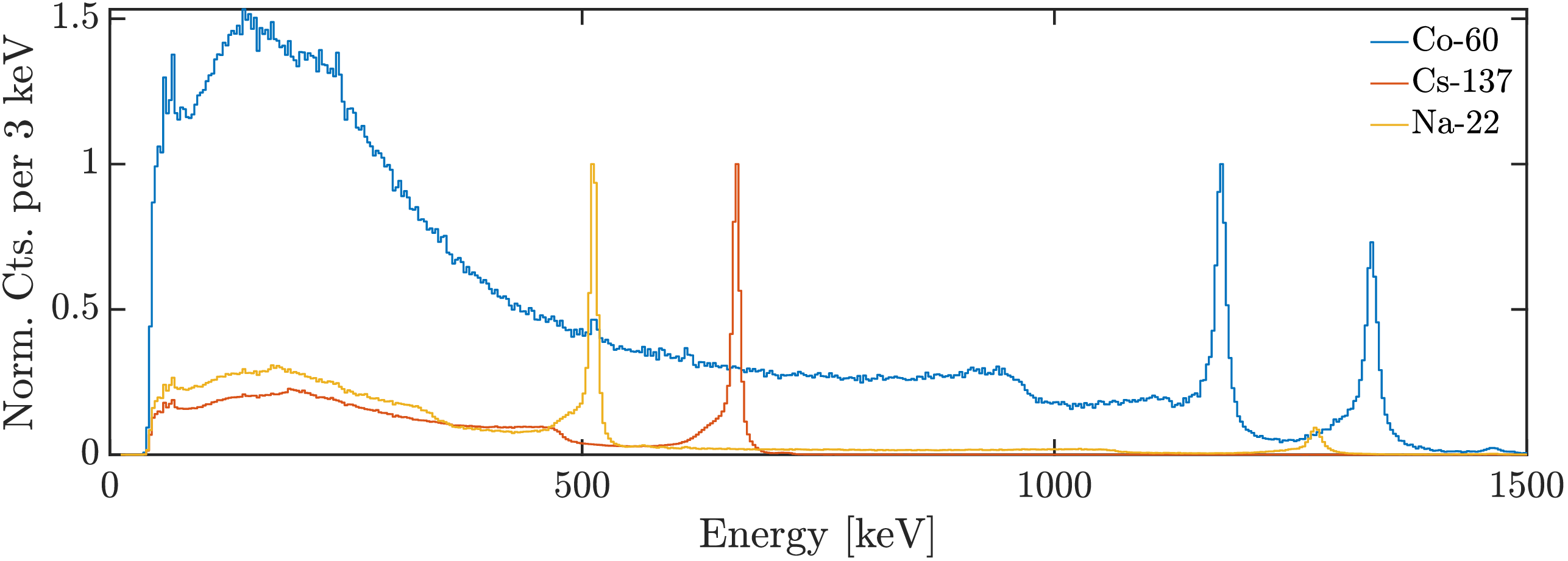}
  \caption{Energy spectra of Co-60, Cs-137, and Na-22 taken by TERI. All pixel events are plotted.}
  \label{fig:highErgSpectra}
\end{figure}

Fig.~\ref{fig:lowErgSpectra} and~\ref{fig:highErgSpectra} plot the spectral response of TERI to different check sources: Am-241, Ba-133, Cs-137, Na-22, and Co-60 (with gamma-ray emission statistics available in~\cite{Helmer_1998}). The energy range for each pixel is $40 \ \mathrm{keV}$ to $3 \ \mathrm{MeV}$. This implies that multi-pixel events can result in the detection of gammas with incident energy higher than $3 \ \mathrm{MeV}$.

\section{Code Aperture Assembly}
\label{sec:ca}

TERI is fitted with a rank 54 pseudorandom-coded aperture mask. We chose a pseudorandom mask, rather than utilize a (M)URA type pattern, as TERI's detectors are sparsely placed. The random noise properties therefore present the appropriate choice rather than a structured mask~\cite{randomMaskGD}. To design the pattern, we create a very large sample of random masks. We then simulated the response by convolving the detector plane's response with the projected pattern and then producing the associated image. We then calculate the mean square error (MSE) of the reconstructed image. This was repeated for source locations uniformly sampled across the field of view. The mask with the smallest overall MSE was chosen. 

\begin{figure}[h!]
     \centering
     \begin{subfigure}[b]{0.45\textwidth}
         \centering
         \subcaption{TERI's Coded Mask Pattern}
         \rotatebox[origin=c]{180}{\includegraphics[trim={0cm 0cm 0cm 0cm}, clip,height=0.7\textwidth]{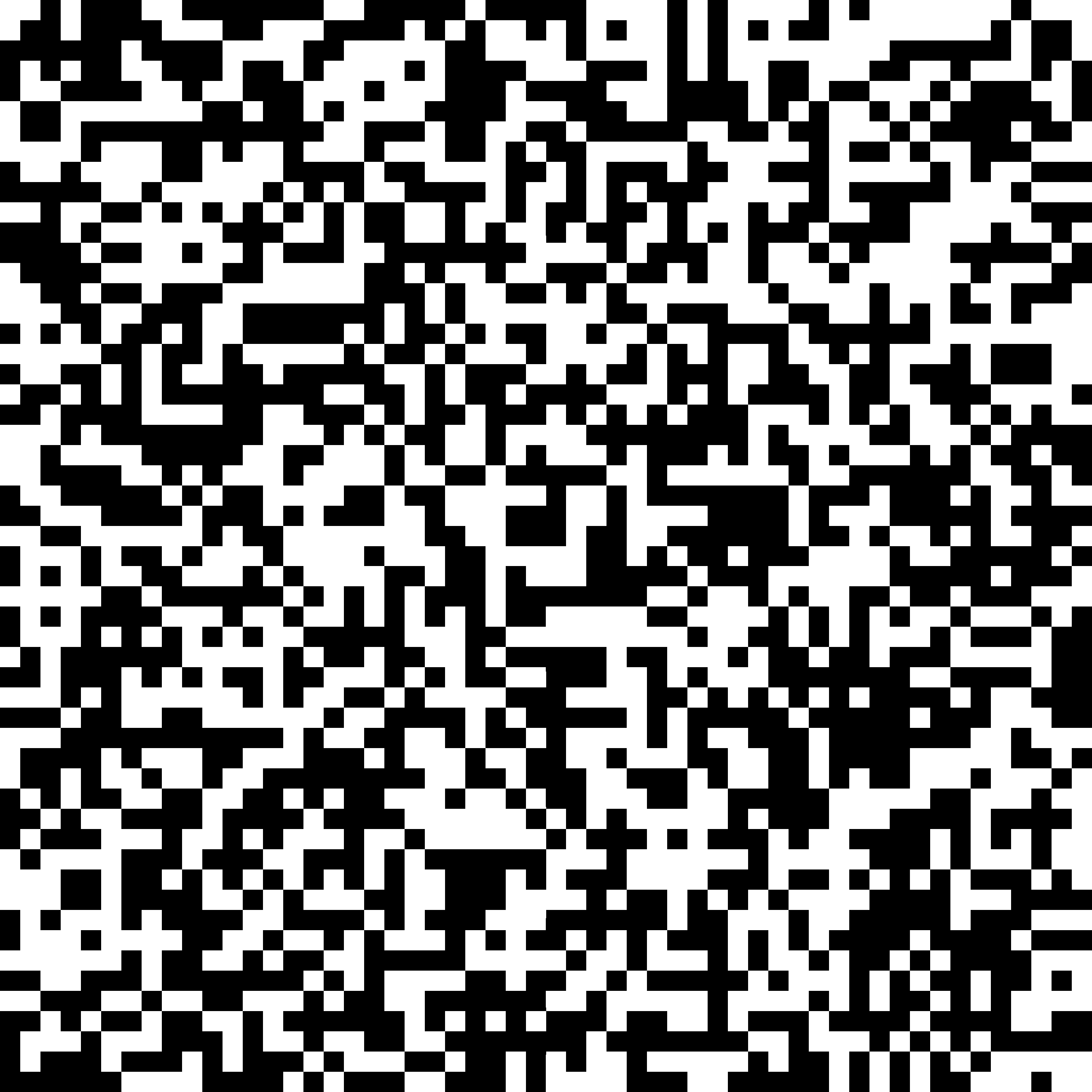}}
         \label{fig:maskPattern}
     \end{subfigure}
     \begin{subfigure}[b]{0.45\textwidth}
         \centering
         \subcaption{TERI Mask}
         \includegraphics[height=0.7\textwidth]{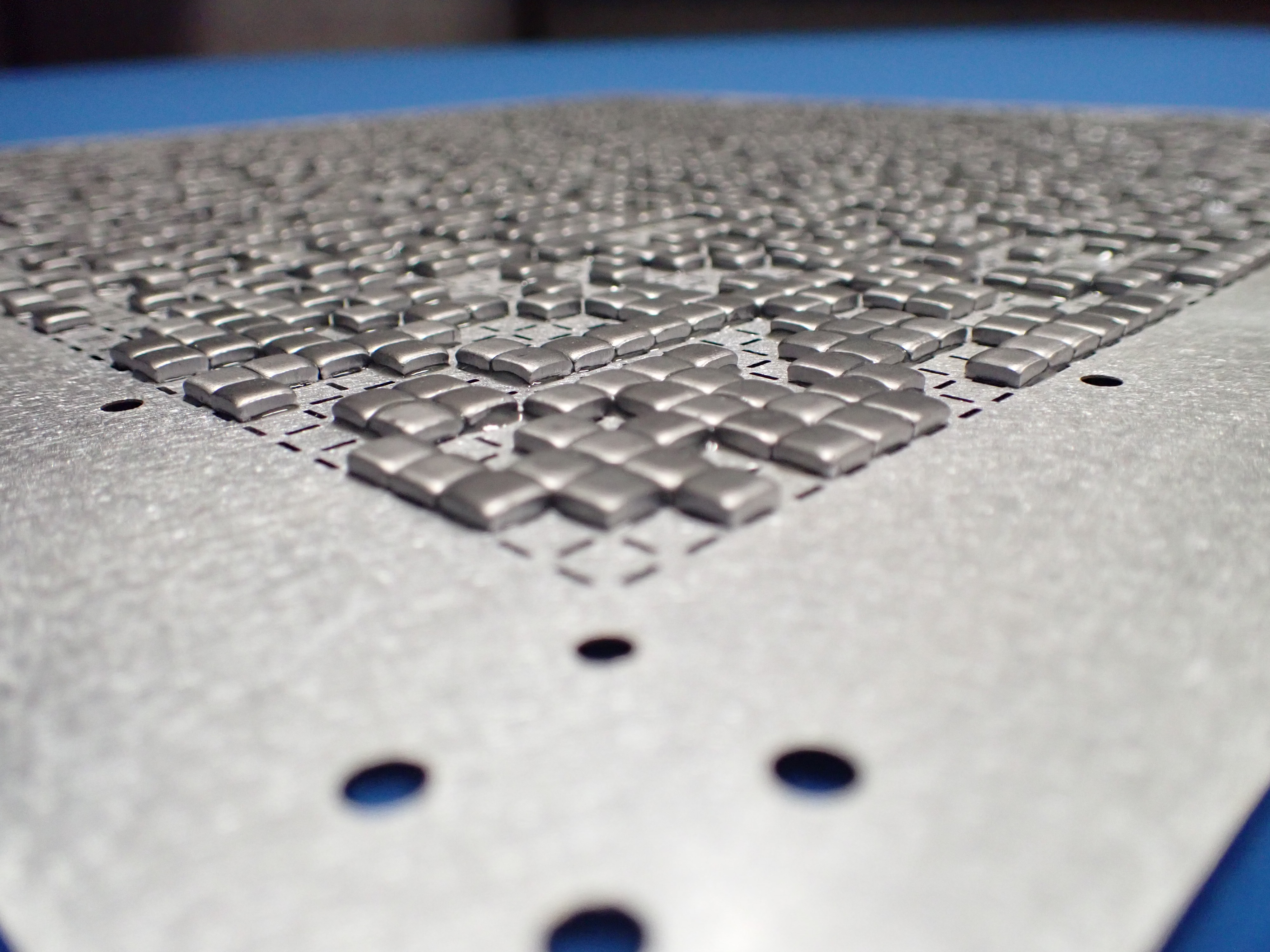}
         \label{fig:codedMaskImage}
     \end{subfigure}

        \caption{Coded Mask of TERI}
        \label{fig:codedMask}
\end{figure}

Fig.~\ref{fig:maskPattern} shows the chosen mask pattern. The mask is made out of 80 mils ($\sim 2 \ \mathrm{mm}$) tantalum square tiles. Each tile is $5.31 \ \mathrm{mm}$ wide, which is 3 times the pitch of the CZT's pixel ($1.77 \ \mathrm{mm}$) to reduce collimation effects. Fig.~\ref{fig:codedMaskImage} shows the assembled mask where each tile is epoxied onto an aluminum panel by hand. The mask-to-detector distance is  6.317 inches with an opening angle of 48.9 deg. The resulting coded field of view is 0.69 steradians.

\subsection{Coded Aperture Imaging}

\begin{figure}[H]
  \centering
  \includegraphics[trim={0cm 0cm 0cm 0cm}, clip, width=0.7\linewidth]{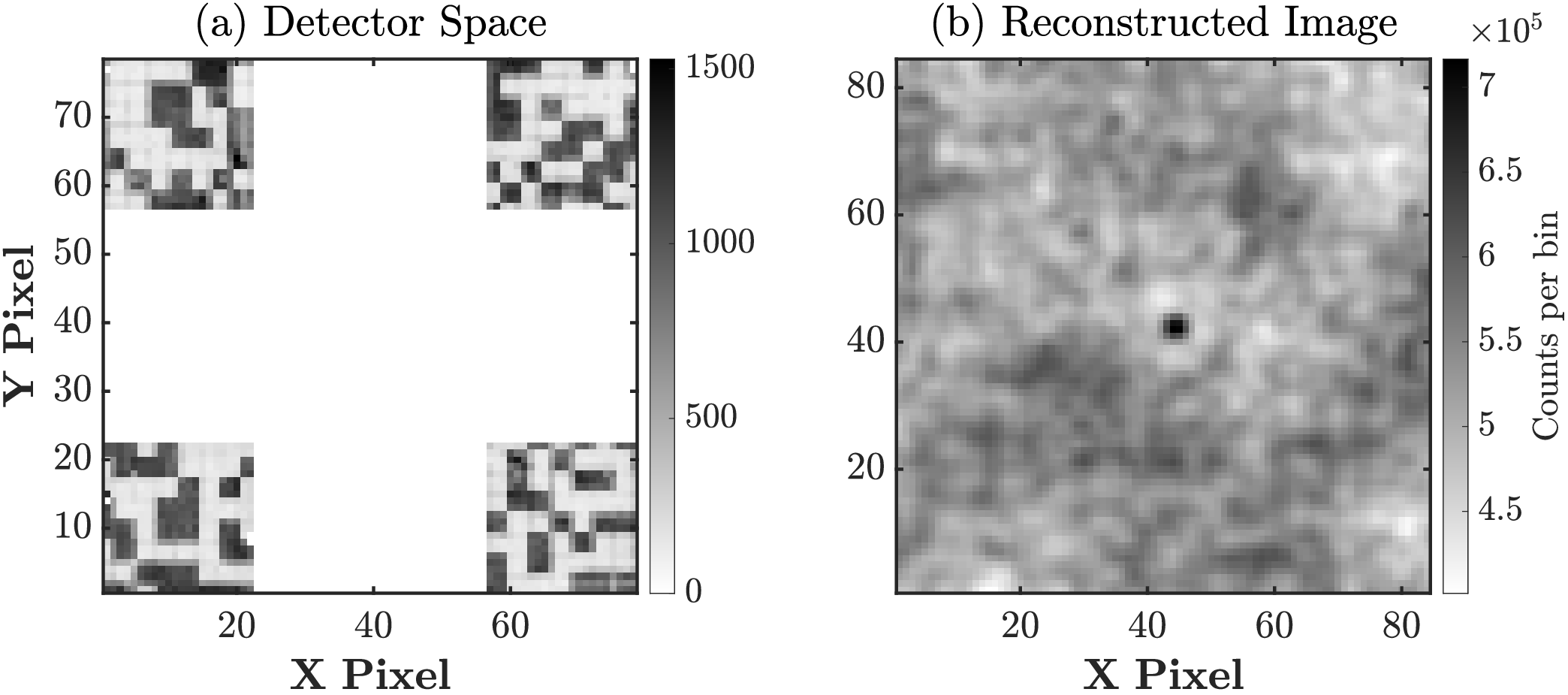}
  \caption{(a) is the recorded pattern by the TERI detectors ($\textbf{Y}$ in Eq.~\ref{eq:x-corr}) when a Co-57 source is placed in the central axis. (b) is the reconstructed image ($\hat{\textbf{X}}$ in Eq.~\ref{eq:x-corr}).}
  \label{fig:firstLight}
\end{figure}

This work uses cross-correlation to reconstruct the images, although more advanced filtering or iterative reconstruction algorithms could be applied. The reconstruction formulation is as follows:

\begin{equation} \label{eq:x-corr}
\hat{\textbf{X}} =  \textbf{D} \otimes \textbf{Y},
\end{equation}

\noindent
with  $\hat{\textbf{X}}$ representing the estimated image, $\textbf{D}$ is the decoding matrix, and $\textbf{Y}$ is the recorded shadowgraph with the measured counts in each pixel representing its elements, $\otimes$ represents the cross-correlation operator. Since TERI utilizes a random mask, the mask is its own decoding function~\cite{accorsi2001design}. Fig.~\ref{fig:firstLight}a displays the recorded pattern by TERI's detector when a Co-57 source is placed in its central axis where a portion of the mask pattern is visible. Using Eq.~\ref{eq:x-corr}, we can reconstruct the image shown in Fig.~\ref{fig:codedMaskImage}. In the image, a point source is visible in the center. We note that there are visible structured artifacts in the background. Those are most likely due to the poor packing fraction of the detector plane and the resulting sparse recording of the projected pattern.

\begin{figure}[H]
  \centering
  \includegraphics[trim={0cm 0cm 0cm 0cm}, clip, width=0.7\linewidth]{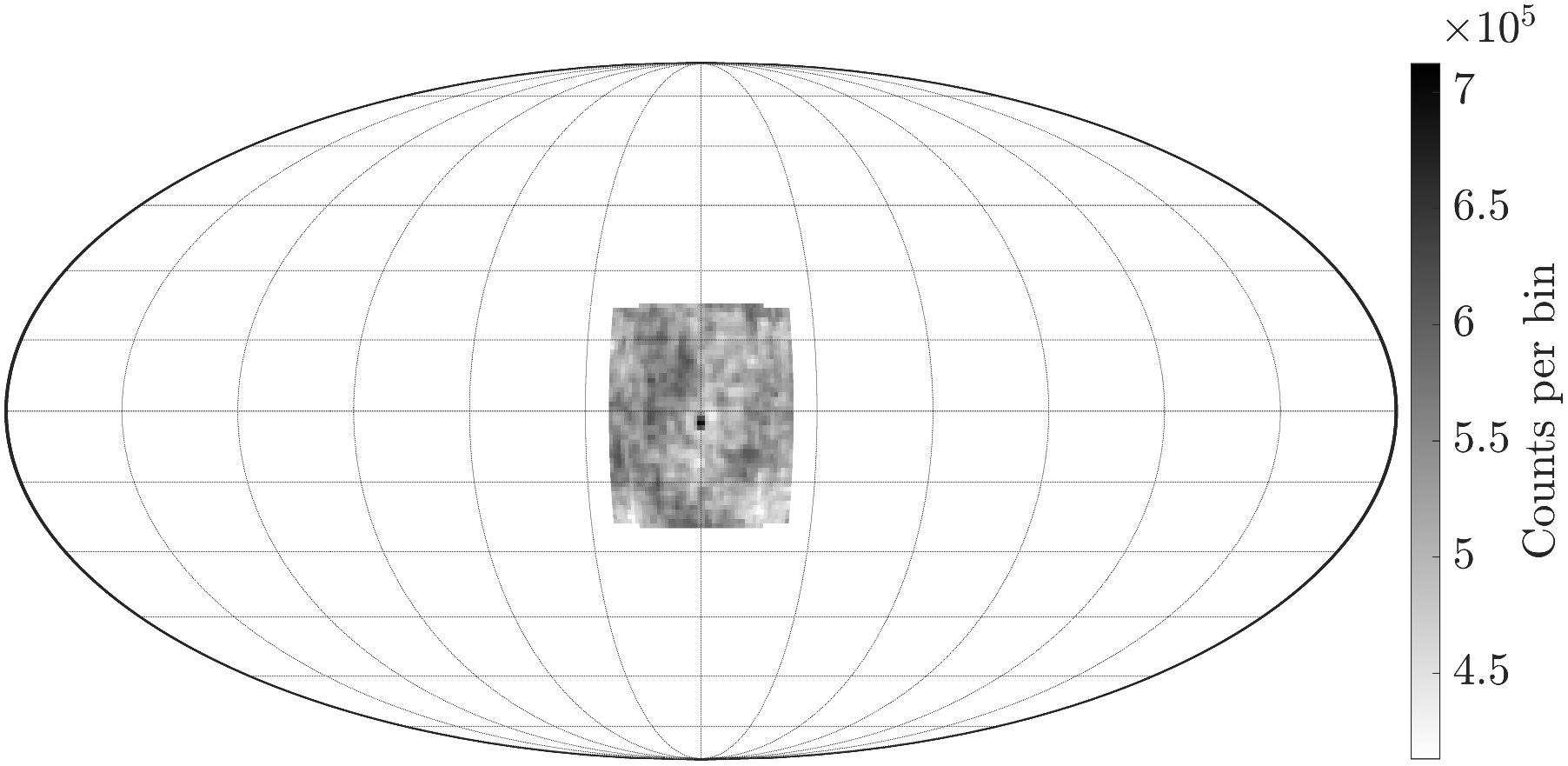}
  \caption{Mollweide projection of a Co-57 source presented in Fig.~\ref{fig:firstLight}b. It is of a Co-57 source placed in the central axis. }
  \label{fig:firstLightMoll}
\end{figure}

Fig.~\ref{fig:firstLightMoll} plots the image presented in Fig.~\ref{fig:firstLight} with a Mollweide projection. This projection displays the field of view of the instrument at any given instance.

\begin{figure}[H]
  \centering
  \includegraphics[trim={0cm 0cm 0cm 0cm}, clip, width=0.7\linewidth]{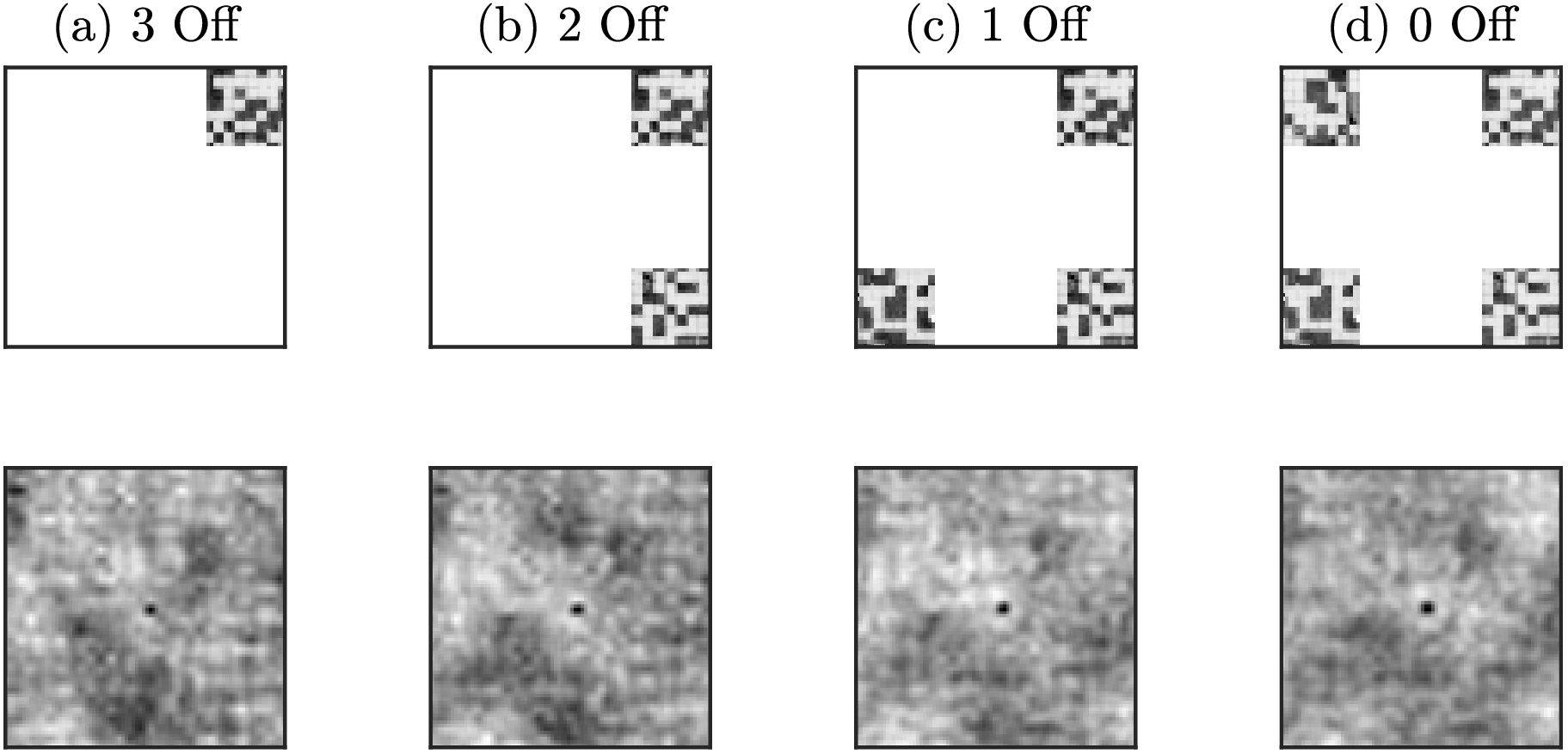}
  \caption{Reconstructed images of a Co-57 source utilizing a different number of detectors. (a) plots the image when 3 detectors are off, (b) 2 are off, (c) 1 is off, while (d) has all detectors operational.}
  \label{fig:turningOff}
\end{figure}

Fig.~\ref{fig:turningOff} shows reconstructed images using TERI where a different number of detectors are turned off. Fig.~\ref{fig:turningOff}a plots the image reconstructed with 3 detectors off and progressively turn on different detectors all the way through Fig.~\ref{fig:turningOff}d where no detectors are turned off. Even with only one detector turned on, TERI can reconstruct a clear image of the source, demonstrating the power of utilizing a random coded mask. As more and more detectors are turned on, the signal-to-noise ratio improves. Note that the background artifacts are changing with the different number of operational detectors.

\section{Compton Imaging}
\label{sec:CI}

Compton imaging is a technique to reconstruct the gamma ray's source position and distribution when the gamma ray's energy is large enough to where Compton scattering is the dominant mode of interaction~\cite{comptonImagingSoft}. Compton imaging was demonstrated using many different families of semiconductors~\cite{SiComptonImaging, HPGeComptonImaging, CZTComptonImaging} and scintillators~\cite{comptel}. 

Briefly, Compton imaging is accomplished if the incident gamma rays undergo Compton scattering followed by a photoelectric absorption. The incident direction of the gamma ray can be determined with the scattering angle between the incident and scattered photons using the Compton scatter equation. The possible locations where the gamma rays can originate are reconstructed as a ring in the image space. As more photons are collected, more rings overlap and eventually converge to the source position and distribution. Compton ring and imaging reconstruction is fully explored in Kierans et alli~\cite{Kierans_2022}. As the TERI detectors are 3D position sensitive, the position of interaction can be determined along with the deposited energy.

At the date of composition of this work, the data pipeline for the full TERI instrument is not yet complete and is still under development. However, we present the gamma-ray imaging performance of one of the four detectors on TERI. We use reconstruction techniques as outlined in Xu~\cite{XuThesis}. This work makes use of two types of reconstruction algorithms. The first is simple backprojection (SBP) which simply projects the Compton rings onto a spherical image space. Since this is known as a biased estimation, we also present maximum likelihood expectation maximization (MLEM) which is an iterative image reconstruction algorithm~\cite{MLEM}. Our MLEM reconstruction techniques follow those outlined in Wilderman et alli~\cite{WildermanMLEM}. We do not implement any advanced iterative stopping criteria and stop the iterative process based on qualitative criteria. The number of interactions per event used in this section range from 2 to 5~\cite{weiyiEIID}. In addition, we apply far-field approximations.

\begin{figure}[H]
  \centering
  \includegraphics[trim={0cm 0cm 0cm 0cm}, clip, width=0.7\linewidth]{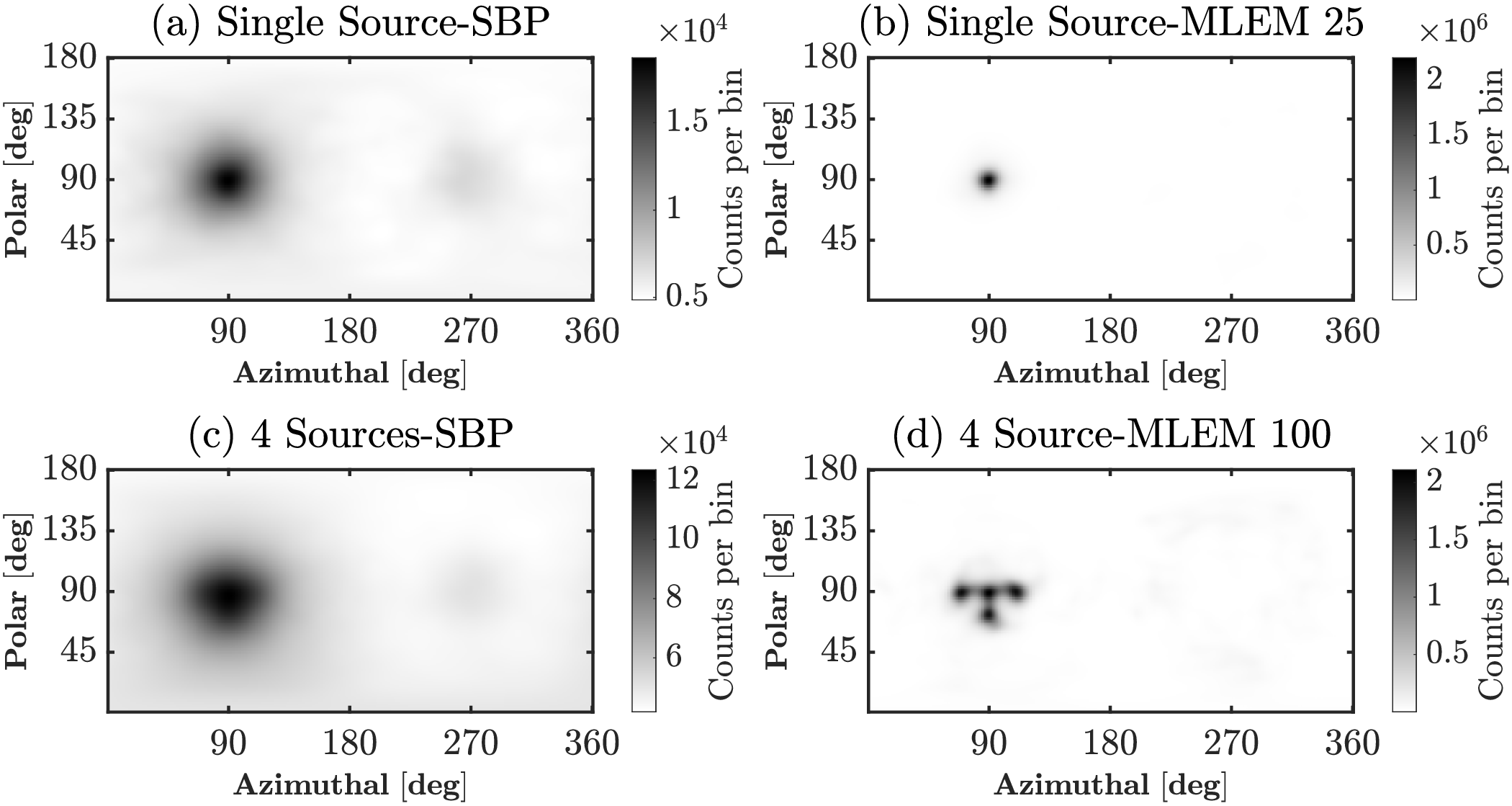}
  \caption{Reconstructed Compton images in equirectangular projections of a Cs-137 source using only 1 TERI detector. The top row presents a single source located at $(90^{\circ}, 90^{\circ})$ reconstructed using (a) simple backprojection and (b) 25 MLEM iterations. The bottom row shows a complex source field where the same Cs-137 source is placed in 4 different location (`T' configuration) and the datasets concatenated to mimic one single observation.}
  \label{fig:compton}
\end{figure}

Figure~\ref{fig:compton}(a) features a SBP reconstruction of a single Cs-137 source placed 36 inches (0.91 meters) directly above the normal of the detector, or $(\mathrm{\theta_{azimuth},\phi_{ polar}})=(90^{\circ}, 90^{\circ})$. The image uses 15k photopeak events and results in a point spread function (PSF) of roughly $45^{\circ}$ FWHM. Figure~\ref{fig:compton}(b) applies MLEM for 25 iterations which results in an FWHM of $<12^{\circ}$. 

We then complexify the source scene by creating a multi-location dataset. Since we do not have multiple sources of the same intensity, we use the same source and place it in multiple locations. We then collect the data for the same runtime at each location. We place the same source in four different locations in a `T' configuration: the center, $20^{\circ}$ to the left, right, and below the center. Using coordinates, this represents $(90^{\circ}, 90^{\circ}), (70^{\circ}, 90^{\circ}), (110^{\circ}, 90^{\circ}), (90^{\circ}, 70^{\circ})$. Figure~\ref{fig:compton}(c) plots the SBP of the 4-source scene using 54k events. The PSF of the hotspot does show some extension relative to the single source scene. We then apply 100 MLEM iterations in Figure~\ref{fig:compton}(d) which clearly unveils the 4-source `T' configuration. As the sources are in the near field and we do not apply any near field correction or advance system response, we observe a relative weakening of the offset sources relative to the center source. Application of an appropriate system response will correct for such effects. This briefly presents TERI's Compton imaging capability for resolving complex sources.

\section{Conclusion}

We developed the cadmium zinc TElluride Radiation Imager (TERI), an instrument that utilizes four large-volume $4 \times 4 \times 1.5 \ \mathrm{cm}^3$ pixelated CdZnTe (CZT). The main objective of TERI is to qualify and characterize the technology for operation in a space environment. Each pixel has an energy range of $40 \ \mathrm{keV}$ up to $3 \ \mathrm{MeV}$ with a resolution of $1.3 \%$ full-width-at-half maximum at $662 \  \mathrm{keV}$. TERI is fitted with a coded aperture which allows for low-energy imaging. In this work, we demonstrated the technologies' coded-aperture imaging and Compton imaging capabilities. TERI is slated for launch to the International Space Station in early 2025 on DoD's STP-H10 mission.

\section{Code and Data Availability}

The authors do not have permission to share data.

\acknowledgments % equivalent to \section*{ACKNOWLEDGMENTS}       
 
We thank the previous researchers whose work on CZT made TERI possible. The authors are particularly grateful to the Department of Defense (DoD) Space Test Program (STP) for providing space launch services and mission operations support for TERI. Daniel Shy appreciates Jacob Levinrad for assisting with mask assembly. Special thanks to Scott Budzien for providing the NRL Payload Operations Control Center. The CdZnTe crystals were provided by the Defense Threat Reduction Agency. Daniel Shy is supported by the U.S. Naval Research Laboratory’s Jerome and Isabella Karle Fellowship. This work is supported by the Office of Naval Research 6.1 funds.

\appendix    %>>>> this command starts appendixes

\section{History of CZT in space based and balloon-borne experiments}
\label{sec:history}

Table~\ref{tab:cztHistory} attempts to plot a comprehensive list of CZT instruments developed specifically for the space and near-space environment. It starts with mostly balloon instruments using thin planar crystals that are on the order of a few millimeters. Over time, the active area of the crystals increased to $4 \times 4 \ \mathrm{cm}^2$ and evolved from planar to pixelated electrode configurations. Nevertheless, the deployed crystals remained under a centimeter of thickness\footnote{We note that in certain applications, like X-ray astrophysics, larger crystals may not be desired.}, which is not ideal for gamma-ray astronomy due to the small interaction probability.

\begin{table}[h!]
  \centering
  \caption{Collection of published balloon and space-based instruments that utilize CZT or CdTe. If two names appear in a single entry that is separated with a colon, the first name corresponds to the name of the instruments while the second name is that of the entire probe. The listed instruments are orbital unless stated otherwise. The `s' shorthands is used to designate slated launch dates. Concepts are limited to known active projects. This list attempts to be comprehensive, however, we acknowledge that we might have missed a few.}
	\begin{tabularx}{\linewidth}{l|c|c|c|c|c}	
	    \toprule
		Name	& Crystal Size 	& Material 	& Electrode Config. & Year Launched & Notes  \\ \hline \hline
		PoRTIA 	& \makecell{$2.54 \times 2.54 \times 0.19 \ \mathrm{cm}^3$ \\ $1.2 \times 1.2 \times 0.5 \ \mathrm{cm}^3$} & CZT & planar & 1995 & balloon~\cite{PoRTIA} \\ \hline
		CfA/CalTech 	& $2.54 \times 2.54 \times 0.19 \ \mathrm{cm}^3$ & CZT & planar & 1995 & balloon~\cite{josh1998} \\ \hline
		HEXIS 	& $1.2 \times 1.2 \times 0.3 \ \mathrm{cm}^3$ & CZT & cross strip/planar & 1997 & balloon~\cite{washUCZT} \\ 	\hline
		InFoc$\mu$s    &  $2.7 \times 2.7 \times 0.2 \ \mathrm{cm}^3$ &   CZT  & pixelated & 2001 & balloon~\cite{InFOCuS} \\  \hline
		ISGRI: Integral    &  $0.4 \times 0.4 \times 0.2 \ \mathrm{cm}^3$ &   CdTe  & planar & 2002 & ~\cite{integral} \\  \hline
		BAT: Swift & $0.4 \times 0.4 \times 0.2 \ \mathrm{cm}^3$ 	& CZT	& planar			& 2005				& ~\cite{swift} \\ \hline
		HEFT & $1.2 \times 2.4 \times 0.2 \ \mathrm{cm}^3$ 	& CZT	& pixelated			& 2005				& balloon~\cite{HEFT} \\ \hline

		GRaND: Dawn	&  $1 \times 1 \times 0.7 \ \mathrm{cm}^3$ 	& CZT 		& 	coplanar	& 2007	&  ~\cite{DAWN_CZT, DAWN_CZT_PERFORMANCE}  \\ \hline
		HEX: Chandrayaan-1    &  $4 \times 4 \times 0.2 \ \mathrm{cm}^3$ &   CZT  & planar & 2008 & ~\cite{Chandrayaan} \\  \hline

		AAUSat-2    &  $1 \times 1 \times 0.4 \ \mathrm{cm}^3$ &   CZT  & pixelated & 2008 & ~\cite{aausat} \\  \hline
		ProtoEXIST    & $1.95 \times 1.95 \times 0.5 \ \mathrm{cm}^3 $ &   CZT  & pixelated & 2009- & balloon~\cite{exist} \\ \hline
		RT-2: Koronas-Foton     & $3.96 \times 3.95 \times 0.5 \ \mathrm{cm}^3 $ &   CZT  & pixelated & 2009 & \cite{rt2} \\ \hline
		NuSTAR    & $2 \times 2 \times 0.2 \ \mathrm{cm}^3 $ &   CZT  & pixelated & 2012 & ~\cite{nustar} \\ \hline
		X(L)-Calibur    &  \makecell{$2 \times 2 \times 0.2 \ \mathrm{cm}^3$ \\ $2 \times 2 \times 0.5 \ \mathrm{cm}^3$ \\ $2 \times 2 \times 0.8 \ \mathrm{cm}^3$} &   CZT  & pixelated & 2014- & balloon~\cite{xlcalibur} \\  \hline
		CZTI: AstroSat    &  $3.9 \times 3.9 \times 0.5 \ \mathrm{cm}^3$ &   CZT  & pixelated & 2015 & ~\cite{czti} \\  \hline
		BeEagleSat    &  $2 \times 2 \times 0.5 \ \mathrm{cm}^3$ &   CZT  & cross strip & 2017 & ~\cite{beeaglesat} \\  \hline

		EPEx    &  $2 \times 2 \times 0.5 \ \mathrm{cm}^3$ &   CZT  & pixelated & 2018 & balloon~\cite{EPEx} \\  \hline

		ASIM    & $2 \times 2 \times 0.5 \ \mathrm{cm}^3$ & CZT & pixelated & 2018 & ~\cite{asim} \\ \hline
		STIX: Solar Orbiter & $1 \times 1 \times 0.1 \ \mathrm{cm}^3$  		& CdTe 		& variable pixelation		&	2020	& ~\cite{stix}    	 \\ \hline
		OrionEagle    & $2 \times 2 \times 1.5 \ \mathrm{cm}^3$  &   CZT  & pixelated & 2021 & balloon~\cite{sara} \\ \hline
		BADG3R    & $2 \times 2 \times 0.6 \ \mathrm{cm}^3$  &   CZT  & cross strip & 2022 & balloon~\cite{badg3r} \\ \hline
				VZLUSAT-2     & $1.4 \times 1.4 \times 0.2 \ \mathrm{cm}^3$  &   CdTe  & pixelated & 2022 & ~\cite{timepix} \\ \hline
		Sharjah-Sat 1     & $2.54 \times 2.54 \times 0.5 \ \mathrm{cm}^3$  &   CZT  & pixelated & 2023 & ~\cite{sharjahSim, iXRD} \\ \hline
		ComPair    & $0.6 \times 0.6 \times 2 \ \mathrm{cm}^3$  &   CZT  & virtual Frisch grid & 2023 & balloon~\cite{comPair, BOLOTNIKOV2024169328}  \\ \hline 
		ECLAIRs: SVOM    &  $0.4 \times 0.4 \times 0.1 \ \mathrm{cm}^3$ &   CdTe  & Schottky-type & 2024 & ~\cite{eclairs} \\

		\hline
		\hline
		MASS-Cube	& $2 \times 2 \times 1.5 \ \mathrm{cm}^3$ & CZT & pixelated & s2024 & ~\cite{MASS} \\ \hline
		AXIS: AEPEX    &  $2 \times 2 \times 0.5 \ \mathrm{cm}^3$ &   CZT  & pixelated & s2024 & ~\cite{AXIS, MARSHALL202066} \\  \hline
		TERI	& $4 \times 4 \times 1.5 \ \mathrm{cm}^3$ & CZT & pixelated & s2025 & this work \\ \hline
		COSI    &  $8 \times 8 \times 1.5 \ \mathrm{cm}^3$ &   HPGe  & cross strip & s2027 & ~\cite{COSI} \\ 
		\hline
		\hline
		Daksha	& $3.9 \times 3.9 \times 0.5 \ \mathrm{cm}^3$ & CZT & pixelated & concept & ~\cite{bhalerao2022daksha} \\ \hline

		ASTENA    &  $0.5 \times 2 \times 2 \ \mathrm{cm}^3$ &   CZT  & strip-drift & concept & ~\cite{ASTENA, Frontera2021} \\ \hline
		MeVCube    &  $2 \times 2 \times 1.5 \ \mathrm{cm}^3$ &   CZT  & pixelated & concept & ~\cite{MeVCube} \\ \hline

		GECCO    &  $0.8 \times 0.8 \times 3.2 \ \mathrm{cm}$ &    CZT  & virtual Frisch grid & concept & ~\cite{GECCO} \\ \hline
		HSP    &  $2 \times 2 \times 0.3 \ \mathrm{cm}^3$ &   CZT  & pixelated & concept & ~\cite{HSP} \\

		\bottomrule		 
	\end{tabularx} 

	\label{tab:cztHistory}
\end{table}

\clearpage

% References
\bibliography{report} % bibliography data in report.bib

\begin{thebibliography}{10}

\bibitem{MeVAstronomy}
Fryer, C.~L. and {et alli}, ``{Catching Element Formation In The Act}.'' arXiv,
  1902.02915 (2019).

\bibitem{swift}
Barthelmy, S.~D., Barbier, L.~M., Cummings, J.~R., Fenimore, E.~E., Gehrels,
  N., Hullinger, D., Krimm, H.~A., Markwardt, C.~B., Palmer, D.~M., Parsons,
  A., Sato, G., Suzuki, M., Takahashi, T., Tashiro, M., and Tueller, J., ``{The
  Burst Alert Telescope (BAT) on the SWIFT Midex Mission},'' {\em Space Science
  Reviews}~{\bf 120},  143--164 (Oct 2005).

\bibitem{czti}
Bhalerao, V., Bhattacharya, D., Vibhute, A., Pawar, P., Rao, A.~R., Hingar,
  M.~K., Khanna, R., Kutty, A. P.~K., Malkar, J.~P., Patil, M.~H., Arora,
  Y.~K., Sinha, S., Priya, P., Samuel, E., Sreekumar, S., Vinod, P., Mithun, N.
  P.~S., Vadawale, S.~V., Vagshette, N., Navalgund, K.~H., Sarma, K.~S.,
  Pandiyan, R., Seetha, S., and Subbarao, K., ``{The Cadmium Zinc Telluride
  Imager on {AstroSat}},'' {\em Journal of Astrophysics and Astronomy}~{\bf 38}
  (jun 2017).

\bibitem{rt2}
Kotoch, T.~B., Nandi, A., Debnath, D., Malkar, J.~P., Rao, A.~R., Hingar,
  M.~K., Madhav, V.~P., Sreekumar, S., and Chakrabarti, S.~K., ``{Instruments
  of RT-2 experiment onboard CORONAS-PHOTON and their test and evaluation II:
  RT-2/CZT payload},'' {\em Experimental Astronomy}~{\bf 29},  27--54 (Feb
  2011).

\bibitem{4x4}
Zhu, Y. and He, Z., ``{Performance of Larger-Volume 40 × 40 × 10- and 40 ×
  40 × 15-mm³ CdZnTe Detectors},'' {\em IEEE Transactions on Nuclear
  Science}~{\bf 68}(2),  250--255 (2021).

\bibitem{cztEfficiency}
Chen, Z., Zhu, Y., and He, Z., ``Intrinsic photopeak efficiency measurement and
  simulation for pixelated cdznte detector,'' {\em Nuclear Instruments and
  Methods in Physics Research Section A: Accelerators, Spectrometers, Detectors
  and Associated Equipment}~{\bf 980},  164501 (2020).

\bibitem{COSI}
Tomsick, J.~A., Boggs, S.~E., Zoglauer, A., Hartmann, D., Ajello, M., Burns,
  E., Fryer, C., Karwin, C., Kierans, C., Lowell, A., Malzac, J., Roberts, J.,
  Saint-Hilaire, P., Shih, A., Siegert, T., Sleator, C., Takahashi, T.,
  Tavecchio, F., Wulf, E., Beechert, J., Gulick, H., Joens, A., Lazar, H.,
  Neights, E., Oliveros, J. C.~M., Matsumoto, S., Melia, T., Yoneda, H., Amman,
  M., Bal, D., von Ballmoos, P., Bates, H., Böttcher, M., Bulgarelli, A.,
  Cavazzuti, E., Chang, H.-K., Chen, C., Chu, C.-Y., Ciabattoni, A.,
  Costamante, L., Dreyer, L., Fioretti, V., Fenu, F., Gallego, S., Ghirlanda,
  G., Grove, E., Huang, C.-Y., Jean, P., Khatiya, N., Knödlseder, J., Krause,
  M., Leising, M., Lewis, T.~R., Lommler, J.~P., Marcotulli, L.,
  Martinez-Castellanos, I., Mittal, S., Negro, M., Nussirat, S.~A., Nakazawa,
  K., Oberlack, U., Palmore, D., Panebianco, G., Parmiggiani, N., Parsotan, T.,
  Pike, S.~N., Rogers, F., Schutte, H., Sheng, Y., Smale, A.~P., Smith, J.,
  Trigg, A., Venters, T., Watanabe, Y., and Zhang, H., ``{The Compton
  Spectrometer and Imager}.'' arXiv, 2308.12362 (2023).

\bibitem{issPic}
``{Flyaround view of {ISS} after undocking}.''
  https://images.nasa.gov/details-s132e012208 (2010).

\bibitem{HeRamoReview}
He, Z., ``{Review of the Shockley–Ramo theorem and its application in
  semiconductor gamma-ray detectors},'' {\em Nuclear Instruments and Methods in
  Physics Research Section A: Accelerators, Spectrometers, Detectors and
  Associated Equipment}~{\bf 463}(1),  250--267 (2001).

\bibitem{KayeThesis}
Kaye, W.~R., {\em {Energy and Position Reconstruction in Pixelated Cadmium Zinc
  Telluride Detectors}}, PhD thesis (2012).
\newblock Copyright - Database copyright ProQuest LLC; ProQuest does not claim
  copyright in the individual underlying works; Last updated - 2023-03-04.

\bibitem{Feng3DCZT}
Zhang, F., He, Z., Xu, D., Knoll, G., Wehe, D., and Berry, J., ``{Improved
  resolution for 3D position sensitive CdZnTe spectrometers},'' in [{\em 2003
  IEEE Nuclear Science Symposium. Conference Record (IEEE Cat.
  No.03CH37515)}{\nolinebreak\hspace{0.1em}]},   {\bf 5},  3356--3360 Vol.5
  (2003).

\bibitem{3dczt}
He, Z., Li, W., Knoll, G., Wehe, D., Berry, J., and Stahle, C., ``{3-D position
  sensitive CdZnTe gamma-ray spectrometers},'' {\em Nuclear Instruments and
  Methods in Physics Research Section A: Accelerators, Spectrometers, Detectors
  and Associated Equipment}~{\bf 422}(1),  173--178 (1999).

\bibitem{HE1999173}
He, Z., Li, W., Knoll, G., Wehe, D., Berry, J., and Stahle, C., ``{3-D position
  sensitive CdZnTe gamma-ray spectrometers},'' {\em Nuclear Instruments and
  Methods in Physics Research Section A: Accelerators, Spectrometers, Detectors
  and Associated Equipment}~{\bf 422}(1),  173--178 (1999).

\bibitem{Helmer_1998}
Helmer, R., ``{Online Spectrum Catalogs for Ge and Si(Li)},'' (1998).

\bibitem{randomMaskGD}
Busboom, A., Elders-Boll, H., and Dieter~Schotten, H., ``{Combinatorial design
  of near-optimum masks for coded aperture imaging},'' in [{\em 1997 IEEE
  International Conference on Acoustics, Speech, and Signal
  Processing}{\nolinebreak\hspace{0.1em}]},   {\bf 4},  2817--2820 vol.4
  (1997).

\bibitem{accorsi2001design}
Accorsi, R., {\em {Design of a near-field coded aperture cameras for
  high-resolution medical and industrial gamma-ray imaging}}, PhD thesis,
  Massachusetts Institute of Technology (2001).

\bibitem{comptonImagingSoft}
Schönfelder, V., Hirner, A., and Schneider, K., ``A telescope for soft gamma
  ray astronomy,'' {\em Nuclear Instruments and Methods}~{\bf 107}(2),
  385--394 (1973).

\bibitem{SiComptonImaging}
Kurfess, J., Johnson, W., Kroeger, R., Novikova, E., Phlips, B., Strickman, M.,
  and Wulf, E., ``{An advanced Compton telescope based on thick,
  position-sensitive solid-state detectors},'' {\em New Astronomy Reviews}~{\bf
  48}(1),  293--298 (2004).
\newblock Astronomy with Radioactivities IV and Filling the Sensitivity Gap in
  MeV Astronomy.

\bibitem{HPGeComptonImaging}
Wulf, E., Phlips, B., Johnson, W., Kroeger, R., Kurfess, J., and Novikova, E.,
  ``{Germanium strip detector Compton telescope using three-dimensional
  readout},'' {\em IEEE Transactions on Nuclear Science}~{\bf 50}(4),
  1182--1189 (2003).

\bibitem{CZTComptonImaging}
Du, Y., He, Z., Knoll, G., Wehe, D., and Li, W., ``{Evaluation of a Compton
  scattering camera using 3-D position sensitive CdZnTe detectors},'' {\em
  Nuclear Instruments and Methods in Physics Research Section A: Accelerators,
  Spectrometers, Detectors and Associated Equipment}~{\bf 457}(1),  203--211
  (2001).

\bibitem{comptel}
Schonfelder, V., Aarts, H., Bennett, K., Deboer, H., Clear, J., Collmar, W.,
  Connors, A., Deerenberg, A., Diehl, R., Von~Dordrecht, A., et~al.,
  ``{Instrument description and performance of the imaging gamma-ray telescope
  COMPTEL aboard the Compton Gamma-Ray Observatory},'' {\em Astrophysical
  Journal Supplement Series}  (1993).

\bibitem{Kierans_2022}
Kierans, C., Takahashi, T., and Kanbach, G.,  [{\em {Compton Telescopes for
  Gamma-Ray Astrophysics}}{\nolinebreak\hspace{0.1em}]},  1–72, Springer
  Nature Singapore (Sept. 2022).

\bibitem{XuThesis}
Xu, D., {\em {Gamma-ray imaging and polarization measurement using
  three-dimensional position-sensitive cadmium zinc telluride detectors}}, PhD
  thesis (2006).

\bibitem{MLEM}
Shepp, L.~A. and Vardi, Y., ``{Maximum Likelihood Reconstruction for Emission
  Tomography},'' {\em IEEE Transactions on Medical Imaging}~{\bf 1}(2),
  113--122 (1982).

\bibitem{WildermanMLEM}
Wilderman, S., Clinthorne, N., Fessler, J., and Rogers, W., ``{List-mode
  maximum likelihood reconstruction of Compton scatter camera images in nuclear
  medicine},'' in [{\em 1998 IEEE Nuclear Science Symposium Conference Record.
  1998 IEEE Nuclear Science Symposium and Medical Imaging Conference (Cat.
  No.98CH36255)}{\nolinebreak\hspace{0.1em}]},   {\bf 3},  1716--1720 vol.3
  (1998).

\bibitem{weiyiEIID}
Wang, W., Wahl, C.~G., Jaworski, J.~M., and He, Z., ``{Maximum-Likelihood
  Deconvolution in the Spatial and Spatial-Energy Domain for Events With Any
  Number of Interactions},'' {\em IEEE Transactions on Nuclear Science}~{\bf
  59}(2),  469--478 (2012).

\bibitem{PoRTIA}
Parsons, A., Barthelmy, S., Bartlett, L., Gehrels, N., Naya, J., Stahle, C.,
  Tueller, J., and Teegarden, B., ``{CdZnTe background measurements at balloon
  altitudes with PoRTIA},'' {\em Nuclear Instruments and Methods in Physics
  Research Section A: Accelerators, Spectrometers, Detectors and Associated
  Equipment}~{\bf 516}(1),  80--95 (2004).

\bibitem{josh1998}
Bloser, P.~F., Grindlay, J.~E., Narita, T., and Harrison, F.~A., ``{CdZnTe
  background measurement at balloon altitudes with an active BGO shield},'' in
  [{\em EUV, X-Ray, and Gamma-Ray Instrumentation for Astronomy
  IX}{\nolinebreak\hspace{0.1em}]},  Siegmund, O. H.~W. and Gummin, M.~A.,
  eds.,  {\bf 3445},  186 -- 196, International Society for Optics and
  Photonics, SPIE (1998).

\bibitem{washUCZT}
{Slavis}, K., {Binns}, W.~R., {Dowkontt}, P., {Epstein}, J., {Hink}, P.,
  {Matteson}, J., {Duttweiler}, F., {Huszar}, G., {Leblanc}, P., {Pelling}, M.,
  {Skelton}, R., and {Stephan}, E., ``{{High Altitude Balloon Flight of CZT
  Detectors for High Energy = X-Ray Astronomy}},'' in [{\em APS April Meeting
  Abstracts}{\nolinebreak\hspace{0.1em}]},  {\em APS Meeting Abstracts},  D3.09
  (Apr. 1998).

\bibitem{InFOCuS}
Baumgartner, W.~H., Tueller, J., Krimm, H., Barthelmy, S.~D., Berendse, F.,
  Ryan, L., Birsa, F.~B., Okajima, T., Kunieda, H., Ogasaka, Y., Tawara, Y.,
  and Tamura, K., ``{InFOCuS hard x-ray telescope: pixellated CZT
  detector/shield performance and flight results},'' in [{\em X-Ray and
  Gamma-Ray Telescopes and Instruments for
  Astronomy}{\nolinebreak\hspace{0.1em}]},  Truemper, J.~E. and Tananbaum,
  H.~D., eds.,  {\bf 4851},  945 -- 956, International Society for Optics and
  Photonics, SPIE (2003).

\bibitem{integral}
{Lebrun, F.}, {Leray, J. P.}, {Lavocat, P.}, {Cr\'etolle, J.}, {Arqu\`es, M.},
  {Blondel, C.}, {Bonnin, C.}, {Bou\`ere, A.}, {Cara, C.}, {Chaleil, T.},
  {Daly, F.}, {Desages, F.}, {Dzitko, H.}, {Horeau, B.}, {Laurent, P.},
  {Limousin, O.}, {Mathy, F.}, {Mauguen, V.}, {Meignier, F.}, {Molini\'e, F.},
  {Poindron, E.}, {Rouger, M.}, {Sauvageon, A.}, and {Tourrette, T.}, ``{ISGRI:
  The INTEGRAL Soft Gamma-Ray Imager *},'' {\em A\&A}~{\bf 411}(1),  L141--L148
  (2003).

\bibitem{HEFT}
Harrison, F.~A., Christensen, F.~E., Craig, W., Hailey, C., Baumgartner, W.,
  Chen, C. M.~H., Chonko, J., Cook, W.~R., Koglin, J., Madsen, K.-K.,
  Pivavoroff, M., Boggs, S., and Smith, D.,  [{\em {Development of the HEFT and
  NuSTAR focusing telescopes}}{\nolinebreak\hspace{0.1em}]},  131--137,
  Springer Netherlands, Dordrecht (2006).

\bibitem{DAWN_CZT}
Prettyman, T., Feldman, W., Fuller, K., Storms, S., Soldner, S., Szeles, C.,
  Ameduri, F., Lawrence, D., Browne, M., and Moss, C., ``{CdZnTe gamma-ray
  spectrometer for orbital planetary missions},'' {\em IEEE Transactions on
  Nuclear Science}~{\bf 49}(4),  1881--1886 (2002).

\bibitem{DAWN_CZT_PERFORMANCE}
Prettyman, T.~H., Feldman, W.~C., McSween, H.~Y., Dingler, R.~D., Enemark,
  D.~C., Patrick, D.~E., Storms, S.~A., Hendricks, J.~S., Morgenthaler, J.~P.,
  Pitman, K.~M., and Reedy, R.~C., ``{Dawn's Gamma Ray and Neutron Detector},''
  {\em Space Science Reviews}~{\bf 163},  371--459 (Dec 2011).

\bibitem{Chandrayaan}
Goswami, J.~N. and Annadurai, M., ``{Chandrayaan-1: India's first planetary
  science mission to the moon},'' {\em Current Science}~{\bf 96}(4),  486--491
  (2009).

\bibitem{aausat}
``{AAUSat}.''
  https://www.eoportal.org/satellite-missions/aausat-2\#rf-communications.

\bibitem{exist}
Hong, J., Grindlay, J., Allen, B., Skinner, G., Barthelmy, S., Gehrels, N.,
  Garson, A., Krawczynski, H., Cook, W., Harrison, F., Natalucci, L., and
  Ubertini, P., ``{The proposed high-energy telescope (HET) for EXIST},'' in
  [{\em Space Telescopes and Instrumentation 2010: Ultraviolet to Gamma
  Ray}{\nolinebreak\hspace{0.1em}]},  Arnaud, M., Murray, S.~S., and Takahashi,
  T., eds.,  {\bf 7732},  77321Y, International Society for Optics and
  Photonics, SPIE (2010).

\bibitem{nustar}
Harrison, F.~A., Craig, W.~W., Christensen, F.~E., Hailey, C.~J., Zhang, W.~W.,
  Boggs, S.~E., Stern, D., Cook, W.~R., Forster, K., Giommi, P., Grefenstette,
  B.~W., Kim, Y., Kitaguchi, T., Koglin, J.~E., Madsen, K.~K., Mao, P.~H.,
  Miyasaka, H., Mori, K., Perri, M., Pivovaroff, M.~J., Puccetti, S., Rana,
  V.~R., Westergaard, N.~J., Willis, J., Zoglauer, A., An, H., Bachetti, M.,
  Barrière, N.~M., Bellm, E.~C., Bhalerao, V., Brejnholt, N.~F., Fuerst, F.,
  Liebe, C.~C., Markwardt, C.~B., Nynka, M., Vogel, J.~K., Walton, D.~J., Wik,
  D.~R., Alexander, D.~M., Cominsky, L.~R., Hornschemeier, A.~E., Hornstrup,
  A., Kaspi, V.~M., Madejski, G.~M., Matt, G., Molendi, S., Smith, D.~M.,
  Tomsick, J.~A., Ajello, M., Ballantyne, D.~R., Baloković, M., Barret, D.,
  Bauer, F.~E., Blandford, R.~D., Brandt, W.~N., Brenneman, L.~W., Chiang, J.,
  Chakrabarty, D., Chenevez, J., Comastri, A., Dufour, F., Elvis, M., Fabian,
  A.~C., Farrah, D., Fryer, C.~L., Gotthelf, E.~V., Grindlay, J.~E., Helfand,
  D.~J., Krivonos, R., Meier, D.~L., Miller, J.~M., Natalucci, L., Ogle, P.,
  Ofek, E.~O., Ptak, A., Reynolds, S.~P., Rigby, J.~R., Tagliaferri, G.,
  Thorsett, S.~E., Treister, E., and Urry, C.~M., ``{THE NUCLEAR SPECTROSCOPIC
  TELESCOPE ARRAY (NuSTAR) HIGH-ENERGY X-RAY MISSION},'' {\em The Astrophysical
  Journal}~{\bf 770},  103 (may 2013).

\bibitem{xlcalibur}
Abarr, Q., Awaki, H., Baring, M., Bose, R., {De Geronimo}, G., Dowkontt, P.,
  Errando, M., Guarino, V., Hattori, K., Hayashida, K., Imazato, F., Ishida,
  M., Iyer, N., Kislat, F., Kiss, M., Kitaguchi, T., Krawczynski, H., Lisalda,
  L., Matake, H., Maeda, Y., Matsumoto, H., Mineta, T., Miyazawa, T., Mizuno,
  T., Okajima, T., Pearce, M., Rauch, B., Ryde, F., Shreves, C., Spooner, S.,
  Stana, T.-A., Takahashi, H., Takeo, M., Tamagawa, T., Tamura, K., Tsunemi,
  H., Uchida, N., Uchida, Y., West, A., Wulf, E., and Yamamoto, R.,
  ``{XL-Calibur – a second-generation balloon-borne hard X-ray polarimetry
  mission},'' {\em Astroparticle Physics}~{\bf 126},  102529 (2021).

\bibitem{beeaglesat}
Kalemci, E., Ümit, E., and Aslan, R., ``{X-ray detector on 2U cubesat
  BeEagleSAT of QB50},'' in [{\em 2013 6th International Conference on Recent
  Advances in Space Technologies (RAST)}{\nolinebreak\hspace{0.1em}]},
  899--902 (2013).

\bibitem{EPEx}
{Cully}, C.~M., {Galts}, D., {Patrick}, M., {Duffin}, C., {Jang}, A.~C.,
  {Pitzel}, J., {Trumpour}, T., {McCarthy}, M., and {Milling}, D.~K., ``{VLF
  and X-ray Instruments for Stratospheric Balloons: ABOVE$^{2}$ and EPEx},'' in
  [{\em AGU Fall Meeting Abstracts}{\nolinebreak\hspace{0.1em}]},   {\bf 2017},
   SH44A--01 (Dec 2017).

\bibitem{asim}
{\O}stgaard, N. et~al., ``{The Modular X- and Gamma-Ray Sensor (MXGS) of the
  ASIM Payload on the International Space Station},'' {\em Space Science
  Reviews}~{\bf 215},  23 (Feb 2019).

\bibitem{stix}
Hayes, L.~A., Musset, S., M{\"u}ller, D., and Krucker, S., ``{The Spectrometer
  Telescope for Imaging X-rays (STIX) on Solar Orbiter},'' (2022).

\bibitem{sara}
Abraham, S., Zhu, Y., Nowicki, S., Bloser, P., Berry, J., Sandoval, B.,
  Lanctot, S., Petryk, M., Deming, J., Klimenko, A., and He, Z., ``{Capability
  demonstration of a 3D CdZnTe detector on a high-altitude balloon flight},''
  {\em Nuclear Instruments and Methods in Physics Research Section A:
  Accelerators, Spectrometers, Detectors and Associated Equipment}~{\bf 1054},
  168413 (2023).

\bibitem{badg3r}
Auricchio, N., Abbene, L., Benassi, G., Bettelli, M., Buttacavoli, A., {Del
  Sordo}, S., Principato, F., {Sarzi Amadè}, N., Stephen, J., Zambelli, N.,
  Zanettini, S., Zappettini, A., and Caroli, E., ``{A CZT 3D imaging
  spectrometer prototype with digital readout for high energy astronomy},''
  {\em Nuclear Instruments and Methods in Physics Research Section A:
  Accelerators, Spectrometers, Detectors and Associated Equipment}~{\bf 1047},
  167869 (2023).

\bibitem{timepix}
Granja, C., Hudec, R., Maršíková, V., Inneman, A., Pína, L., Doubravova,
  D., Matej, Z., Daniel, V., and Oberta, P., ``{Directional-Sensitive
  X-ray/Gamma-ray Imager on Board the VZLUSAT-2 CubeSat for Wide Field-of-View
  Observation of GRBs in Low Earth Orbit},'' {\em Universe}~{\bf 8}(4) (2022).

\bibitem{sharjahSim}
Altıngün, A.~M., Kalemci, E., and Öztaban, E., ``{A simulation study for the
  expected performance of Sharjah-Sat-1 payload improved X-Ray Detector (iXRD)
  in the orbital background radiation},'' {\em Experimental Astronomy}~{\bf
  56},  117–140 (Jan. 2023).

\bibitem{iXRD}
Kalemci, E., Alt{\i}ng{\"u}n, A.~M., Bozkurt, A., Aslan, A.~R.,
  Yal{\c{c}}{\i}n, R., G{\"o}kalp, K., Veziro{\u{g}}lu, K., Fernini, I.,
  Manousakis, A., Ya{\c{s}}ar, A., Diba, M., Karabulut, B., {\c{C}}atal, E.,
  and {\"O}ztekin, O., ``{The Improved X-ray Detector (iXRD) on Sharjah-Sat-1,
  design principles, tests and ground calibration},'' {\em Experimental
  Astronomy}~{\bf 56},  99--116 (Aug 2023).

\bibitem{comPair}
Shy, D., Kierans, C., Cannady, N., Caputo, R., Griffin, S., Grove, J.~E., Hays,
  E., Kong, E., Kirschner, N., Liceaga-Indart, I., McEnery, J., Mitchell, J.,
  Moiseev, A.~A., Parker, L., Perkins, J.~S., Phlips, B., Sasaki, M.,
  Schoenwald, A.~J., Sleator, C., Smith, J., Smith, L.~D., Wasti, S., Woolf,
  R., Wulf, E., and Zajczyk, A., ``{Development of the ComPair gamma-ray
  telescope prototype},'' in [{\em Space Telescopes and Instrumentation 2022:
  Ultraviolet to Gamma Ray}{\nolinebreak\hspace{0.1em}]},  den Herder,
  J.-W.~A., Nikzad, S., and Nakazawa, K., eds.,  {\bf 12181},  121812G,
  International Society for Optics and Photonics, SPIE (2022).

\bibitem{BOLOTNIKOV2024169328}
Bolotnikov, A., Carini, G., Dellapenna, A., Deptuch, G., Fried, J., Herrmann,
  S., Laassiri, M., Lee, W., Maj, P., Moiseev, A., Pinaroli, G., Sasaki, M.,
  Smith, L., Tamura, E., and Yates, E., ``{3x3 array module of 8×8×32 mm3
  position-sensitive virtual Frisch-grid CdZnTe detectors for imaging and
  spectroscopy of cosmic gamma-rays},'' {\em Nuclear Instruments and Methods in
  Physics Research Section A: Accelerators, Spectrometers, Detectors and
  Associated Equipment} ,  169328 (2024).

\bibitem{eclairs}
Schanne, S., ``{The ECLAIRs telescope onboard the SVOM mission for gamma‐ray
  burst studies},'' {\em AIP Conference Proceedings}~{\bf 1000}(1),  581--584
  (2008).

\bibitem{MASS}
Zhu, J., Zheng, X., Feng, H., Zeng, M., Huang, C.-Y., Hsiang, J.-Y., Chang,
  H.-K., Li, H., Chang, H., Pan, X., Ma, G., Wu, Q., Li, Y., Bai, X., Ge, M.,
  Ji, L., Li, J., Shen, Y., Wang, W., Wang, X., Zhang, B., and Zhang, J., ``Mev
  astrophysical spectroscopic surveyor (mass): a compton telescope mission
  concept,'' {\em Experimental Astronomy}~{\bf 57},  2 (Feb 2024).

\bibitem{AXIS}
Berland, G.~D., Marshall, R.~A., Martin, C., Buescher, J., Kohnert, R.~A.,
  Boyajian, S., Cully, C.~M., McCarthy, M.~P., and Xu, W., ``{The atmospheric
  X-ray imaging spectrometer (AXIS) instrument: Quantifying energetic particle
  precipitation through bremsstrahlung X-ray imaging},'' {\em Review of
  Scientific Instruments}~{\bf 94},  023103 (02 2023).

\bibitem{MARSHALL202066}
Marshall, R.~A., Xu, W., Woods, T., Cully, C., Jaynes, A., Randall, C., Baker,
  D., McCarthy, M., Spence, H.~E., Berland, G., Wold, A., and Davis, E., ``{The
  AEPEX mission: Imaging energetic particle precipitation in the atmosphere
  through its bremsstrahlung X-ray signatures},'' {\em Advances in Space
  Research}~{\bf 66}(1),  66--82 (2020).
\newblock Advances in Small Satellites for Space Science.

\bibitem{bhalerao2022daksha}
Bhalerao, V., Vadawale, S., Tendulkar, S., Bhattacharya, D., Rana, V., Adalja,
  H. K.~L., Belatikar, H., Bhaganagare, M., Dewangan, G., Ghodgaonkar, A.,
  Goyal, S.~K., Gunasekaran, S., Guruprasad, P.~J., Koyande, J.~G., Kulkarni,
  S., Kutty, A., Ladiya, T., Marla, D., Mate, S., Mithun, N. P.~S., Mote, R.,
  Narang, S., Nema, A., Nimbalkar, S., Pai, A., Palit, S., Patel, A., Patel,
  J., Pradeep, P., Ramachandran, P., Saiguhan, B. S.~B., Saraogi, D., Sawant,
  D., Shanmugam, M., Sharma, P., Shetye, A., Singh, S., Singh, N., Singhal, A.,
  Sreekumar, S., Sridhar, S., Srinivasan, R., Tallur, S., Tiwari, N.~K.,
  Vadladi, A.~L., Vaishnava, C.~S., Vishwakarma, S., and Waratkar, G.,
  ``{Daksha: On Alert for High Energy Transients},'' (2022).

\bibitem{ASTENA}
Moita, M., Ferro, L., Caroli, E., Virgilli, E., Frontera, F., Stephen, J.~B.,
  Curado Da~Silva, R.~M., Maia, J.~M., and Del~Sordo, S., ``{ASTENA’s
  Polarimetric Prospects},'' in [{\em 2021 IEEE Nuclear Science Symposium and
  Medical Imaging Conference (NSS/MIC)}{\nolinebreak\hspace{0.1em}]},   1--7
  (2021).

\bibitem{Frontera2021}
Frontera, F., Virgilli, E., Guidorzi, C., Rosati, P., Diehl, R., Siegert, T.,
  Fryer, C., Amati, L., Auricchio, N., Campana, R., Caroli, E., Fuschino, F.,
  Labanti, C., Orlandini, M., Pian, E., Stephen, J.~B., Del~Sordo, S.,
  Budtz-Jorgensen, C., Kuvvetli, I., Brandt, S., da~Silva, R. M.~C., Laurent,
  P., Bozzo, E., Mazzali, P., and Valle, M.~D., ``{Understanding the origin of
  the positron annihilation line and the physics of supernova explosions},''
  {\em Experimental Astronomy}~{\bf 51},  1175--1202 (Jun 2021).

\bibitem{MeVCube}
Lucchetta, G., Ackermann, M., Berge, D., and Bühler, R., ``{Introducing the
  {MeVCube} concept: a {CubeSat} for {MeV} observations},'' {\em Journal of
  Cosmology and Astroparticle Physics}~{\bf 2022},  013 (aug 2022).

\bibitem{GECCO}
Orlando, E., Bottacini, E., Moiseev, A., Bodaghee, A., Collmar, W., Ensslin,
  T., Moskalenko, I.~V., Negro, M., Profumo, S., Digel, S.~W., Thompson, D.~J.,
  Baring, M.~G., Bolotnikov, A., Cannady, N., Carini, G.~A., Eberle, V.,
  Grenier, I.~A., Harding, A.~K., Hartmann, D., Herrmann, S., Kerr, M.,
  Krivonos, R., Laurent, P., Longo, F., Morselli, A., Philips, B., Sasaki, M.,
  Shawhan, P., Shy, D., Skinner, G., Smith, L.~D., Stecker, F.~W., Strong, A.,
  Sturner, S., Tomsick, J.~A., Wadiasingh, Z., Woolf, R.~S., Yates, E., Ziock,
  K.-P., and Zoglauer, A., ``{Exploring the MeV sky with a combined coded mask
  and Compton telescope: the Galactic Explorer with a Coded aperture mask
  Compton telescope (GECCO)},'' {\em Journal of Cosmology and Astroparticle
  Physics}~{\bf 2022},  036 (jul 2022).

\bibitem{HSP}
{Grindlay}, J., {Allen}, B., {Hong}, J., {Violette}, D., {Barthelmy}, S.,
  {Lien}, A., {Elvis}, M., {Steiner}, J., {Tomsick}, J., {Markwardt}, C., and
  {Miyasaka}, H., ``{{High Resolution Energetic X-ray Imager SmallSat
  Pathfinder (HSP) to enable 4piXIO}},'' in [{\em American Astronomical Society
  Meeting Abstracts \#235}{\nolinebreak\hspace{0.1em}]},  {\em American
  Astronomical Society Meeting Abstracts} {\bf 235},  159.04 (Jan. 2020).

\end{thebibliography}
\bibliographystyle{spiebib} % makes bibtex use spiebib.bst

\end{document}